\documentclass[a4paper,fleqn,usenatbib,linenumbers]{mnras}

\usepackage{mathtools}

\usepackage[T1]{fontenc}
\usepackage{ae,aecompl}
\usepackage{tablefootnote}
\usepackage{pdflscape}
\usepackage{hyperref}
\usepackage[dvipsnames]{xcolor}

\usepackage{graphicx}	
\usepackage{amsmath}	
\usepackage{amssymb}	
\usepackage{subfig}
\usepackage{float}
\usepackage{multirow}
\usepackage{units}
\usepackage{placeins}
\usepackage{appendix}
\usepackage{xcolor}

\title[Disc-Jet coupling in BH-XRBs]{Revisiting Disc-Jet Coupling in Black Hole X-ray Binaries: On the Nature of Disc Dynamics and Jet Velocity}

\author[Das Choudhury et al.]{Sreetama Das Choudhury$^{1}$\thanks{E-mail: d.sreetama@iitg.ac.in}, Bhuvana G. R.$^{2}$\thanks{E-mail: bhuvanahebbar@gmail.com}, Santabrata Das$^{1}$\thanks{E-mail: sbdas@iitg.ac.in} and Anuj Nandi$^{3}$\thanks{E-mail: anuj@ursc.gov.in}\\
$^{1}$Department of Physics, Indian Institute of Technology Guwahati, Guwahati, 781039, India.\\
$^{2}$Institute for Advanced Study, Gifu University, 1-1 Yanagido, Gifu 501-1193, Japan.\\
$^{3}$Space Astronomy Group, ISITE Campus, U. R. Rao Satellite Center, Outer Ring Road, Marathahalli, Bengaluru, 560037, India.\\
}

\date{Accepted XXX. Received YYY; in original form ZZZ}

\pubyear{2023}

\begin{document}
\label{firstpage}
\pagerange{\pageref{firstpage}--\pageref{lastpage}}
\maketitle

\begin{abstract}
    We perform a comprehensive wide-band ($3-100$ keV) spectro-temporal analysis of thirteen outbursting black hole X-ray binaries (BH-XRBs), utilizing data (quasi)simultaneous with radio observations to unravel the complex disc-jet connection. \textit{RXTE} observations are analyzed for XTE J$1859+226$, GX $339-4$ (2002, 2006, and 2010 outbursts), 4U $1543-47$, H$1743-322$ (2003 and 2009 outbursts), XTE J$1550-564$, XTE J$1752-223$, XTE J$1650-500$, Swift J$1753.5-0127$, XTE J$1748-288$, and GRO J$1655-40$. For Swift J$1727.8-1613$ and MAXI J$1535-571$, we utilize \textit{HXMT} data, while both \textit{AstroSat} and \textit{HXMT} observations are analyzed for Swift J$1658.2-4242$. Type-C QPOs observed in harder states (LHS, HIMS; $F_{nth} \ge 0.4$) exhibit positive lag for low-inclination sources ($i<50^{\circ}$), whereas it generally exhibits negative lag for high-inclination sources ($i>60^{\circ}$), except XTE J$1550-564$, Swift J$1727.8-1613$, H$1743-322$ (2003 outburst) and GRO J$1655-40$. Notably, type-A QPOs exhibit negative lags ($\sim 1-10$ ms) regardless of source inclination, while type-B QPOs show positive lags in low-inclination sources, and both positive and negative lags ($\sim 1-15$ ms) in high-inclination sources, typically occurring in SIMS ($F_{nth}\lesssim0.45$). Systematic appearance of type-A QPOs preceding radio flares in several sources suggests that type-A QPOs indicate telltale signs of jet ejection, while type-B QPOs are closely linked with radio flares ($i.e.$, transient jets). Present findings suggest the corona evolves from a radially extended to a vertically elongated structure during the type-C to type-B transition via type-A QPOs, with type-B QPOs linked to radially compact or vertically extended coronal geometries, resembling jet ejection. The strong radio–X-ray luminosity correlation seems to provide compelling evidence of accretion-powered jets. Finally, we find that jets in SIMS are moderately relativistic in nature with velocities $\gtrsim 0.3-0.8\,c$ in BH-XRBs under consideration.
        
\end{abstract}


\section{Introduction} 

Galactic black hole X-ray binaries (BH-XRBs) are considered as the ideal cosmic laboratories to understand the accretion-ejections mechanism \cite[and references therein]{Mirabel-Rodriguez-1994,Fender-etal-2004,Fender-etal-2009,Belloni-etal-2011,Miller-etal-2012,Radhika-Nandi-2014,Radhika-etal-2016,Bright-etal-2020,Espinasse-etal-2020,Nandi-etal-2024}. Indeed, the imprint of disc-jet connections are encoded in the spectro-temporal features of the outbursting BH-XRBs. During an outbursting phase, BH-XRBs exhibit outflow activities in the form of jets, which after being ejected, expands adiabatically and emits synchrotron radiations in infrared and radio. A correlation between X-ray and radio lightcurves is observed in many BH-XRBs \cite[]{Mirabel-Rodriguez-1998}, where the radio emission is observed after a delay \cite[]{Mirabel-Rodriguez-1998,Corbel-etal-2005,Russell-etal-2019,Homan-etal-2020,Espinasse-etal-2020,Monageng-etal-2021} with respect to X-ray emission. Hence, these cosmic entities offer an excellent opportunity for exploring the accretion-ejection processes in strong gravity environment around black holes.

BH-XRBs are categorized either as persistent \cite[]{Chen-etal-1997,Remillard-McClintock-2006,Tetarenko-etal-2016,Corral-Santana-etal-2016} or transient \cite[]{Tanaka-Shibazaki-1996,Remillard-McClintock-2006,Sreehari-etal-2018} based on their overall characteristics and therefore, these sources are expected to undergo evolution driven by their accretion dynamics. It is worth mentioning that the evolution of accretion discs of outbursting BH-XRBs are studied through the Hardness Intensity Diagram (HID) \cite[also referred as Q-diagram;][]{Maccaroni-Coppi-2003,Belloni-2004,Homan-Belloni-2005,Belloni-etal-2005,Remillard-McClintock-2006,Motta-etal-2009,Nandi-etal-2012,Aneesha-etal-2019,Sreehari-Nandi-2021,Nandi-etal-2024}, which typically traces out all four spectral states, namely Low Hard State (LHS), Hard Intermediate State (HIMS), Soft Intermediate State (SIMS) and High Soft State (HSS) \cite[]{Homan-etal-2001,Fender-etal-2004,Belloni-etal-2005, Nandi-etal-2012, Radhika-Nandi-2014, Sreehari-etal-2018, Nandi-etal-2018,Blessy-etal-2020,Nandi-etal-2024}. Interestingly, different spectral states are associated with the various types of QPOs, which are generally classified into three categories, such as type-C, type-B, and type-A \cite[and references therein]{Wijnands-etal-1999, Homan-etal-2001, Remillard-etal-2002, Homan-Belloni-2005, Ingram-etal-2009, Nandi-etal-2012, Radhika-etal-2018}. Type-C QPOs are characterized by frequencies $\nu_{\rm QPO} \sim 0.1-30$ Hz, quality factor ${\rm Q} \sim 10$ and rms amplitudes ${\rm rms}_{\rm QPO}\% \sim 5-20$, and are observed during LHS, HIMS, and sometimes in SIMS \cite[]{Casella-etal-2004}. Their power density spectra (PDS) typically show flat-top noise at low frequencies, followed by red noise at higher frequencies, often accompanied by sub-harmonics and harmonics. Type-B QPOs are observed with $\nu_{\rm QPO} \sim 1-7$ Hz \cite[see also][]{Gao-etal-2014}, ${\rm Q} \sim 3-6$ and rms$_{\rm QPO}\% \sim 3-5$. Their PDS is dominated by red noise and may occasionally show harmonics features. On contrary, type-A QPOs are weak and broad with $\nu_{\rm QPO} \sim 5-8$ Hz, ${\rm Q} \le 3$, and rms$_{\rm QPO}\% \le 4$. Their PDS is dominated by red noise without any detection of harmonics. Type-A and Type-B QPOs are typically observed during the SIMS. What is more is that \cite{Casella-etal-2004, Casella-etal-2005} introduced additional QPO classifications, such as type-C* and type-B cathedral QPOs, which are later reported by \cite{Rodriguez-etal-2011} and \cite{Radhika-Nandi-2014}.

Usually, LHS and HIMS exhibit type-C QPOs along with compact radio emission characterized by the flat or inverted spectrum. In these spectral states, the X-ray energy spectra are mostly dominated by the Comptonized high energy radiations over the thermal disc emission, which often best described by the power-law distribution \cite[and references therein]{Chakrabarti-Titarchuk-1995,Zdziarski-etal-1996,Mandal-Chakrabarti-2005,Done-etal-2007, Motta-etal-2009,Iyer-etal-2015,Bhuvana-etal2023,Aneesha-etal2024,Banerjee-etal-2024}. In the outbursting evolutionary track, HIMS is followed by the SIMS, which usually remains softer compared to the HIMS with a relatively stronger dominance of thermal emission. Quite often, SIMS exhibits type-A and type-B QPOs \cite[]{Belloni-etal-2005,Remillard-McClintock-2006,Fender-etal-2009,Motta-etal-2010,Radhika-Nandi-2014,Gao-etal-2014,Radhika-etal-2016,Homan-etal-2020,Harikrishna-Sriram-2022,Liu-etal-2022,Peirano-etal-2023, Zhang-etal-2023, Ma-etal-2024}, and optically thin transient relativistic radio jets are occasionally observed \cite[]{Fender-etal-2009,McClintock-etal-2009,Miller-etal-2012,Carotuneto-etal-2021,Wood-etal-2021,Wood-etal-2024}. Finally, the source is evolved to HSS with strong thermal emission and relatively weak signature of Comptonized emission, where the jet activity is suppressed completely and QPO signature disappears in the power spectra \cite[]{Motta-etal-2010,Miller-etal-2012,Radhika-Nandi-2014,Radhika-etal-2016,Zhang-etal-2023}.

Generally, the ejection of jets occurs during the transition from the HIMS to SIMS, as denoted by the jet line in HID \cite[]{Fender-etal-2004}. Moreover, type-B or type-A QPOs are observed concurrent with radio flares, suggesting the occurrence of jet ejection \cite[]{Fender-etal-2004,Fender-etal-2009,Varniere-etal-2012,Radhika-etal-2016,Homan-etal-2020, Kylafis-etal-2020, Sriram-etal-2021, Harikrishna-Sriram-2022, Zhang-etal-2023}. Earlier work of \cite{Blandford-Znajek-1977} suggests that the rotational energy of the spinning BH can be extracted and subsequently utilized to power the jets. Indeed, this appealing mechanism indicates that spin of the black hole plays pivotal role in jet activities. Meanwhile, a positive correlation between black hole spin and jet luminosity is observed in several BH-XRBs \cite[]{Steiner-etal-2012,Narayan-McClintock-2013,McClintock-etal2014}. However, conflicting claims are reported indicating weak correlation between source spin and the observed jet power \cite[]{Fender-etal-2010,Russell-etal-2013,Aktar-etal-2015,Aktar-etal-2017}. Interestingly, in spite of having a plethora of X-ray and radio observations, the origin as well as the intrinsic mechanisms responsible for powering the jets still remain inconclusive.

Outbursting BH-XRBs exhibit two types of radio jets, namely compact/steady and transient \cite[]{Fender-etal-2004}. The quiescent state and LHS are characterized by low bulk velocity jets (Lorentz factor $\Gamma \leq 2$) \cite[]{Fender-etal-1999,Miller-etal-2012}. As X-ray luminosity increases, jet power also increases, showing a non-linear correlation $L_\text{radio} \propto L_{\text{X}}^{0.6-0.7}$ during low hard state \cite[]{Gallo-etal-2003,Corbel_etal-2003,Corbel-etal-2013,Kylafis-etal-2023}. With increasing X-ray luminosity, jets in outbursting sources gradually become unstable followed by intense radio emissions indicating the ejection of a faster-moving radio lobe \cite[and references therein]{Fender-etal-2009}. When colliding with previously ejected jet material, internal shocks are generated, resulting in the production of a bright, optically thin relativistic jets \cite[]{Miller-etal-2012}. Beside this, an alternative scenario is also suggested, where the excess thermal gradient force across the centrifugally supported shocks diverts a part of the inflowing matter in the form of bipolar jets \cite[]{Chakraborti-1999,Das-etal-2001,Kumar-Chattopadhyay2013,Aktar-etal-2015,Aktar-etal-2017,Das-etal-2022,Joshi-etal-2022}. These jets are collimated and accelerated by the radiations emanated from the disc \cite[]{Chattopadhyay-etal-2005}. This conjecture is supported by the reduction in hard flux component during SIMS, suggesting the evacuation of corona in launching the bipolar jets \cite[]{Vadawale-etal-2001,Nandi-etal-2001, Radhika-Nandi-2014, Nandi-etal-2018}. Indeed, optically thick jets in hard states appear continuous, steady, and mildly relativistic ($\beta \sim 0.1$) \cite[]{Dhawan-etal-2000}, while optically thin jets in SIMS resolve into plasmoids moving at relativistic speeds ($\beta \sim 0.9$) \cite[]{Mirabel-Rodriguez-1999}. In BH-XRBs, the jet velocity is commonly determined by analyzing resolved images that depict the proper motion of ejected material \cite[]{Mirabel-Rodriguez-1994, Corbel-etal-2005, Miller-etal-2012}. It is worth mentioning that assessing the relativistic nature of transient jets remains challenging for numerous BH-XRBs due to the non-availability of resolved images of radio ejecta \cite[]{Fender-etal-2004}.

Meanwhile, numerous efforts were made to understand the dynamics of the corona as well as the origin of QPOs through the time/phase lag studies \cite[]{Reig-etal-2000,Belloni-etal-2005, Dutta-etal-2016,Eijnden-etal-2017,Zhang-etal-2023}. In general, it is argued that positive (hard) lag occurs due to inverse Comptonization of soft photons in the corona \cite[]{Payne-1980, Miyamoto-etal-1988, Reig-etal-2000, Dutta-etal-2016,Mendez-etal-2022}, whereas negative (soft) lag is caused due to Compton down-scattering \cite[]{Reig-etal-2000}, reverberation \cite[]{Uttley-etal-2014, Karpouzas-etal-2020, Zhang-etal-2023} and gravitational bending \cite[]{Chatterjee-etal-2017}. Meanwhile, \cite{Gao-etal-2014} argued that the type-B QPO observed in GX $339-4$ is associated with the inverse-Comptonization of soft photons in the corona. Similarly, \cite{Chatterjee-etal-2020} reported an increase in positive time lag with QPO frequency during the rising phase of the outburst, suggesting that outflows may possibly contribute to positive lags. \cite{Munoz-Darias-etal-2010} observed unusually large time lags ($\sim 0.1$ s) which appear to be resulted from inverse-Comptonization occurring in the outflows.

Notably, numerous efforts were put forward to examine the origin of type-B QPOs in the realm of vertically extended corona. \cite{Belloni-etal-2020} suggested a jet-like corona in MAXI J1348$-$630 based on observed positive lags, whereas \cite{Homan-etal-2020} reported a $\sim 2.5$ hour delay between type-B QPOs and jet ejection. \cite{Stevens-Uttley-2016,Kylafis-etal-2020} proposed the precessing jet base as the hard X-ray source, while \cite{Mendez-etal-2022} argued for vertically elongated corona near radio flares in GRS 1915$+$105. In contrast, \cite{Zhang-etal-2023} attributed type-B QPOs involving the precession of an elongated corona, where hard photons fall back to the disc and are reprocessed resulting in negative lags. Furthermore, \cite{Liu-etal-2022} argued that type-B QPOs are linked to weaker jets than type-C QPOs.

Previous studies have examined the connections between the type of QPOs, source inclination angles, and time lags for selected BH-XRBs \cite[]{Dutta-etal-2016,Eijnden-etal-2017}. Incidentally, the role of source inclination in understanding the lag behavior associated with type-A QPOs remains inconclusive. In addition, the connection between the lag characteristics of different QPO types and the jet ejection also remains elusive. Moreover, there has been lack of attempts to estimate jet velocities by correlating radio and X-ray properties during SIMS of BH-XRBs.

Keeping the above considerations in mind, we carry out an in-depth spectro-temporal analysis of X-ray data along with the available radio observations to understand the underlying physical processes active in BH-XRBs. Since jets are believed to originate from the inner accretion disc \cite[]{Miller-etal-2012, Mendez-etal-2022, Kylafis-etal-2020}, investigating the evolution of the corona characteristics during state transitions provides insight into the disc-jet connection. In this work, we explore the dynamics of the inner accretion disc through wide-band ($3-100$ keV) spectral and timing analyses of thirteen BH-XRBs, namely XTE J$1859+226$, GX $339-4$ (2002, 2006, and 2010 outbursts), 4U $1543-47$, H$1743-322$ (2003 and 2009 outbursts), XTE J$1550-564$, XTE J$1752-223$, XTE J$1650-500$, Swift J$1753.5-0127$, XTE J$1748-288$, GRO J$1655-40$, Swift J$1727.8-1613$, MAXI J$1535-571$ and Swift J$1658.2-4242$, using \textit{RXTE}, \textit{HXMT} and \textit{AstroSat} observations. Source selection is done based on the good coverage in radio and X-ray observations, and the wide range of inclination angles ($\sim35^\circ$ \text{to} $\sim75^\circ$). With this, we investigate the spectro-temporal properties of these sources and using both X-ray and radio observations, we estimate the jet velocities during radio flares. Finally, we attempt to explore the spectro-temporal correlations and examine the role of accretion dynamics in powering jets.

The paper is organized as follows: In Section \ref{sec:section2}, we briefly describe the data reduction procedure both in X-rays and radio. In Section \ref{sec:section3}, we discuss the modelling of PDS, time lag calculations, modelling the energy spectra and spin measurement using continuum fitting method. We present our results in Section \ref{sec:section4}. In Section \ref{sec:section5}, we estimate the source spin and jet velocity during the radio flares. Finally, we discuss the implications of our results in Section \ref{sec:section6} and conclude.

\section{Observation and Data Reduction} 
\label{sec:section2}

In this work, we examine the inherent disc-jet coupling in several Galactic outbursting BH-XRBs observed with \textit{RXTE}\footnote[1]{\url{https://heasarc.gsfc.nasa.gov/cgi-bin/W3Browse/w3browse.pl}}, \textit{HXMT}\footnote[2]{\url{https://archive.hxmt.cn/proposal}} and \textit{AstroSat} \cite[]{Agarwal-2016}. In order to facilitate this, we make use of (quasi-)simultaneous radio observations available as well. We choose sixteen outbursts from thirteen such BH-XRBs, namely XTE J$1859+226$,  H$1743-322$ (2003 and 2009 outbursts), GX $339-4$ (2002, 2007 and 2010 outburst), 4U $1543-47$, XTE J$1752-223$, XTE J$1650-500$,  Swift J$1753.5-0127$, XTE J$1550-564$, XTE J$1748-288$, GRO J$1655-40$, Swift J$1727.8-1613$, MAXI J$1535-571$, Swift J$1658.2-4242$, respectively, which have adequate coverage in X-ray as well as radio. Among them, XTE J1859$+$226, XTE J1752$-$223 and Swift J1727.8$-$1613 showed multiple X-ray and radio peaks in a single outburst during 1999, 2009 and 2023, respectively, whereas remaining sources under considerations exhibited single(multiple) outburst(s) till date.

\subsection{X-ray Data}
\label{subsec:subsection2.1}

We use thirteen \textit{RXTE} and three \textit{HXMT} and one \textit{AstroSat} archival observations of outbursting BH-XRBs. For XTE J1859$+$226, the outburst was observed in $1999-2000$ \cite[]{Smith-1999} which was continued for $163$ days (MJD $51460$ to MJD $51626$). GX 339$-$4 exhibited several outbursts. In this study, we focus on the $2002-2003$ outburst of GX~339$-$4 \cite[]{Smith-etal-2002}, which lasted for $477$ days, using data from MJD~$52288.6$ to MJD~$52765.6$ for our analysis. We also consider the $2007$ outburst \cite[]{Soldi-etal-2007}, with data spanning from MJD~$54102.31$ to MJD~$54256.58$, consisting of $154$ days and the $2010$ outburst \cite[]{Yamaoka-etal-2010}, covering MJD~$55208.48$ to MJD~$55260.07$, comprising of 51 days. For $2002$ outburst of 4U 1543$-$47 \cite[]{Miller-Remillard-2002}, we analyse $73$ days (MJD $52442.8$ to MJD $52515.5$) of pointed observations. For H1743$-$322, the $2009$ outburst \cite[]{Miller-etal-2009, Kalemci-etal-2005} persisted for $73$ days, spanning from MJD~$54980.4$ to MJD~$55053.8$. In addition, we consider the $2003$ outburst of H1743$-$322 \cite[]{Remillard-etal-2003}, specifically selecting observations that are pertinent to time lag analysis and jet velocity estimations. We consider $1998$ outburst of XTE J1550$-$564 \cite[]{Rutledge-etal-1998} and analyze data for $254$ days (MJD $51065$ to MJD $51318$). For 2009 outburst of XTE J1752$-$223 \cite[]{Markwardt-etal-2009}, we analyze observations from MJD $55130$ to MJD $55414$, with a duration of $284$ days. In case of XTE J1650$-$500, we consider the 2002 outburst \cite[]{Tomsick-etal-2002} from MJD $52158$ to MJD $52416$ for a duration of $258$ days. For Swift J1753.5$-$0127, we analyze 2005 outburst \cite[]{Morgan-etal-2005} that lasts for $150$ days from MJD $53553$ to MJD $53703$. For Swift J1727.8$-$1613, we consider $42$ days of observations (MJD $60181$ to $60222$) from $2023$ outburst \cite[]{Negoro-etal-2023}. For XTE~J$1748$–288, we focus on the $1998$ outburst \cite[]{Smith-etal-1998}, considering observations spanning from MJD~$50968.83$ to MJD~$51082.14$. In the case of GRO~J$1655$–40, we examine the $2005$ outburst \cite[]{Markwardt-etal-2005}, utilizing data recorded between MJD~$53421.36$ and MJD~$53685.20$, with $263$ days of observations. For MAXI~J$1535$–571, we analyze the $2017$ outburst \cite[]{Negoro-etal-2017}, covering the period from MJD~$58002.45$ to MJD~$58177.83$, with 18 pointed observations. Finally, for Swift~J$1658.2$–4242, we consider observations from its $2018$ outburst \cite[]{Grebenev-etal-2018}, spanning MJD~$58169.98$ to MJD~$58206.77$ and consisting of 7 pointed observations by \textit{AstroSat} and 25 pointed observations by \textit{Insight/HXMT}.

We use Binned mode, Event mode and Good Xenon data of \textit{RXTE} for our timing analysis. In order to extract the lightcurves in the energy range $2-15$ keV and $15-30$ keV, we utilize standard \texttt{FTOOLS} tasks, namely \texttt{saextrct} for Binned mode data and \texttt{seextrct} for Event mode and Good Xenon data \cite[see][and references therein]{Nandi-etal-2012,Radhika-Nandi-2014}. Subsequently, we use \texttt{lcmath} task to merge two lightcurves to obtain the resultant lightcurve in $2-30$ keV energy range in obtaining the power density spectrum.

For wide-band ($3-100$ keV) spectral modelling, we utilize \texttt{Standard2} data products from PCA and extract the source spectrum, background spectrum and response file using standard \texttt{FTOOLS} tasks, such as \texttt{seextrct}, \texttt{pcabackest} and \texttt{pcarsp}, respectively. In order to generate the background spectrum for PCA data, we use the bright background model\footnote[3]{\url{https://heasarc.gsfc.nasa.gov/docs/xte/pca\_bkg\_epoch.html}} and SAA passage history from SAA webpage\footnote[4]{\url{https://heasarc.gsfc.nasa.gov/docs/xte/pca_history.html}}. In case of HEXTE data, we separate the source and background raw FITS file from Science data product using \texttt{hxtback} task for all sources. We extract the source and background spectra, perform the deadtime correction and finally generate the response file using \texttt{seextrct}, \texttt{hxtdead} and \texttt{hxtrsp} tasks, respectively.

The \textit{Hard X-ray Modulation Telescope} also known as \textit{Insight-HXMT} \cite[]{Zhang-etal-2014} comprises of three instruments: low-energy X-ray telescope (LE: $1-15$ keV; \cite{Chen-etal-2020}), medium-energy X-ray telescope (ME: $5-30$ keV; \cite{Cao-etal-2020}) and high-energ X-ray telescope (HE: $20-250$ keV \cite[]{Liu-etal-2020}). We utilize data from all three instruments for our spectro-temporal analysis of Swift J1727.8$-$1613, MAXI J$1535-571$ and Swift J$1658.2-4242$. However, data beyond MJD $60222$ for Swift J$1727.8-1613$ is not available due to Sun constrain \cite[]{Yu-etal-2024}. We use the \texttt{hpipeline} under \textit{Insight-HXMT} Data Analysis Software (\texttt{HXMTDAS}) version v2.06 to reduce the data. Data extraction is done using the criteria recommended by the \textit{Insight-HXMT} team, $i.e.$, pointing offset $\le0.04^{\circ}$, Earth elevation angle $\ge10^{\circ}$, geomagnetic rigidity cut-off value $\ge8$ GV and finally, $300$ s before and after SAA. Similar to data from \textit{RXTE}, the lightcurve is generated with time resolution $8$ ms from energy range $2-30$ keV. 

\textit{AstroSat}, India’s first dedicated multiwavelength space observatory launched in 2015, is equipped with a suite of X-ray instruments, including the Soft X-ray Telescope (\textit{SXT}) \cite[]{Singh-etal-2017}, the Large Area X-ray Proportional Counter (\textit{LAXPC}) \cite[]{Yadav-etal-2016, Antia-etal-2017}, and the Cadmium Zinc Telluride Imager (\textit{CZTI}) \cite[]{Yadav-etal-2016}. For the present study, we employed the \textit{LAXPC10} and \textit{LAXPC20} detectors for timing analysis of Swift~J$1658.2$–4242, and used only \textit{LAXPC20} for spectral analysis. The \textit{LAXPC} instrument is sensitive to X-rays in the $3-80$~keV energy range \cite[]{Yadav-etal-2016, Agarwal-etal-2017, Antia-etal-2017}. Data reduction from level-1 to level-2 was carried out using \texttt{LAXPCSoftv3.4.4}\footnote[5]{\url{https://www.tifr.res.in/~AstroSat_LAXPC/LAXPCSoft.html}}, the latest software package released on June 21, 2023 \cite[]{Antia-etal-2017}.

\subsection{Radio Data}
\label{subsec:subsection2.2}

We compile the radio flux data for BH-XRBs from the existing literature. In particular, observations of the source XTE J1859$+$226 across a range of radio frequencies are obtained from \cite{Brocksopp-etal-2002}, and due to dense coverage  within the $1.4-1.66$ GHz band, we focus solely on radio fluxes within this frequency range for our analysis. The radio flux for XTE J1550$-$564 is obtained from \cite{Hannikainen-etal-2001} at $8.6$ GHz. Radio flux data for 4U 1543$-$47 is obtained from \cite{Park-etal-2004,Kalemci-etal-2005}. Radio flux values from the 2009 outburst of H1743$-$322 are extracted from \cite{Miller-etal-2012}. We use radio flux for XTE J1752$-$223 from \cite{Brocksopp-etal-2013}. For XTE J1650$-$500, radio flux values are taken from \cite{Corbel-etal-2004}. For GX~339$-$4, we utilize radio fluxes from its 2002, 2007 and 2010 outbursts as reported by \cite{Gallo-etal-2003} and \cite{Corbel-etal-2013, Islam-etal-2018}. Radio observations of XTE~J1748$-$288 are taken from \cite{Brocksopp-etal-2007}, and for GRO~J1655$-$40, we adopt the flux values from \cite{Shaposhnikov-etal-2007}. MAXI~J1535$-$571 and Swift~J1658.2$-$4242 radio fluxes are obtained from \cite{Russell-etal-2019} and \cite{Bogensberger-etal-2020}, respectively. Radio flux for Swift J1753.5$-$0127 is obtained from \cite{Soleri-etal-2010} and lastly, the radio observations in Swift J1727.8$-$1613 are taken from \cite{Peters-etal-2023}. Moreover, flux values for the 2003 outburst of H1743$-$322 are explicitly used for jet velocity calculation \cite[]{McClintock-etal-2009}. Note that to estimate the jet velocity, we normalize the radio flux values for $5$ GHz as,
\begin{equation}
\text{F}(5 ~\text{GHz})=\text{F}(\nu)\times \left(\frac{5~\text{GHz}}{\nu}\right)^\alpha,
\label{eq:eqn1}
\end{equation}
where $\alpha$ denotes the spectral index and $\nu$ is the frequency of observation.

\section{Analysis and modelling}
\label{sec:section3}

We revisit the archival data of thirteen outbursting BH-XRBs and carry out the spectro-temporal analysis of these sources to extract key timing and spectral parameters. This comprehensive approach enables us to explore the underlying disc-jet connection, likely driven by the complex accretion processes occurring around these sources.

\subsection{Timing Analysis}
\label{subsec:subsection3.1}

We generate the power density spectrum (PDS) from $2-30$ keV lightcurve in rms space using \texttt{powspec} task and apply $\text{norm}=-2$ to remove the Poisson noise. The QPO feature is modelled in \texttt{XSPEC} environment (\texttt{v12.14.0}) using multiple \texttt{Lorentzian} and \texttt{powerlaw} model components. With this, we estimate total rms in the frequency range $0.01-62.5$ Hz and QPO rms as rms$_{\rm QPO}\%=\sqrt{P\times\Delta\nu}\times 100$, where $P$ denotes the power in the unit of ${\rm rms}^2 {\rm ~Hz}^{-1}$ and $\Delta \nu$ is the width of the frequency bin \cite[]{Reimann-1867,Belloni-Hasinger-1990,Radhika-etal-2018,Bhuvana-etal-2021}. 

Time lag analysis is performed by generating lightcurves in the energy ranges $2-6$ keV and $6-15$ keV, with the $2-6$ keV lightcurve serving as the reference energy band. The simultaneous photon counts of the lightcurves in these two energy bands at time $t_{\rm k}$ are denoted as $x_{\rm a}(k)$ and $x_{\rm b}(k)$. Their Fourier transforms are given by \cite[]{VanderKlis-1989, MajumderP-etal2024},
\begin{align}
    X_{a}(j)=\sum_{ k=0}^{N-1}x_{a}(k)\exp(2\pi i\nu_{j}t_{k}),\\
    X_{b}(j)=\sum_{ k=0}^{N-1}x_{b}(k)\exp(2\pi i\nu_{ j}t_{k}),
\end{align}
where $x_{a}(k)\equiv x_a(t_{k})$ and $X_{a}(j) \equiv X_{a}(\nu_j)$. Here, $k$ and $j$ represent the time and frequency bins, such that $k \in [0, N-1]$ and $j \in [-N/2,N/2-1]$. In addition, $N$ indicates the total number of bins of the time series having time length $T$ and time step $\delta {\rm t}=T/N$. Thus, $t_{k}=kT/N$ refers time in $k_{\rm th}$ bin and $\nu_{j}=j/T$ denotes frequency at $j_{\rm th}$ bin of the equidistant time and frequency series data.
		
Cross spectrum between these two Fourier transforms is given as $C(j)=X^*_{a}(j)X_{b}(j)$ \cite[]{Nowak-etal-1999}, where $X^*_{a}(j)$ is the Fourier transform of $2-6$ keV light curve and $X_{b}(j)$ is the Fourier transform of $6-15$ keV lightcurve at frequency $\nu_j$. The phase lag is given by the argument/position angle of $C(j)$ in the complex plane and is calculated as,
\begin{equation}
\phi=arg[C(j)].
\end{equation}
Finally, the time lag at the QPO frequency $\nu_j$ is estimated as,
\begin{equation}
\delta t(j)=\frac{C(j)}{2\pi \nu_j}.
\end{equation}
The time lag is computed as the average over the frequency range $\nu_{\text{QPO}} \pm \text{FWHM}$ \cite[]{Reig-etal-2000}, where FWHM denotes the full width at half maximum of the QPO. To calculate the time lag energy spectrum, the $2-6$ keV band is selected as the reference, with $6-10$ keV, $10-15$ keV, $16-20$ keV, $20-24$ keV, and $24-30$ keV as the subject bands. The time lag is calculated using the Stingray \texttt{v.1.1.2} module \cite[]{Huppenkothen-etal-2019} in the \texttt{astropy} package of Python.

\subsection{Spectral Analysis}
\label{subsec:subsection3.2}

We perform wide-band ($3-100$ keV) spectral modeling using data from the PCA ($3-40$ keV) and HEXTE ($20-100$ keV) instruments of the \textit{RXTE} mission. We use data from PCU-2/PCA and Cluster-A/HEXTE for all observations of all sources under considerations except for $2009$ outburst of H1743$-$322, where Cluster-B data are analyzed. For \textit{Insight/HXMT} observations, we use LE ($2-10$ keV), ME ($8-35$ keV) and HE ($28-120$ keV) for broadband spectral analysis. In case of \textit{AstroSat}, we use \textit{LAXPC20} ($3-60$ keV) for spectral analysis. Throughout the analysis, $1\%$ systematic error is considered for spectral modeling using \texttt{XSPEC} (\texttt{version $12.14.1$d}).

In order to carry out the spectral modeling, we first adopt a model combination \texttt{tbabs$\times$(diskbb+cutoffpl)$\times$constant}. The model component \texttt{tbabs} \cite[]{Wilm-etal-2000} is used to explain the interstellar absorption. The thermal and non-thermal emissions of the accretion disc are taken care by the model components \texttt{diskbb} \cite[]{Makishima-etal-1986} and \texttt{cutoffpl}. We use this model combination for the spectral fitting of one BH-XRB (as an example XTE J1859$+$226) that yields reasonable fit with $\chi^2_{red} \sim 0.7-1.4$ for most of the observations. We obtain $E_{\rm cut} \sim 50 - 90$ keV in LHS and HIMS. In SIMS and HSS, $E_{\rm cut}$ could not be directly constrained, and hence, we perform spectral modelling by fixing $E_{\rm cut}$ at discrete values in $5$ keV intervals. The optimal $E_{\rm cut}$ is identified as the value that yielded the least variations in $\chi^2_{\rm red}$. Based on this analysis, $E_{\rm cut}$ is fixed in the range $\sim 50 - 60$ keV for SIMS and $\sim 10$ keV for HSS. However, \texttt{cutoffpl} being a phenomenological model failed to quantify certain physical parameters, such as scattering fraction and electron temperature. 
 
Hence, we adopt the convolution model \texttt{thcomp} \cite[]{Zdziarski-etal-2020} that successfully explains the spectral characteristics at high energies for BH-XRBs under considerations. Accordingly, we use a model combination \texttt{tbabs$\times$(thcomp$\otimes$diskbb)$\times$constant}. To account for the reflection and absorption features, we use \texttt{Gaussian} in the energy range $6-7$ keV, and \texttt{smedge} \cite[]{Ebisawa-etal-1994} based on its usefulness in previous studies \cite[]{Sobczak-etal-1999,Tomsick-etal-2000,Yamaoka-etal-2012,Aneesha-etal-2019,Dong-etal-2020}. Note that \texttt{Gaussian} model suffices for all BH-XRBs except XTE J1859$+$226, where both \texttt{Gaussian} and \texttt{smedge} are required to obtain best fit \cite[]{Radhika-Nandi-2014,Nandi-etal-2018}. The spectral modeling yields spectral index $\Gamma$ in the range $1.6-1.8$ during LHS for all sources. Notably, we are unable to constrain electron temperature $kT_e$ for number of observations during SIMS, and hence, we freeze $kT_e$ to obtain the best fit. In doing so, we perform multiple spectral fits by fixing $kT_e$ at discrete intervals of $5$ keV as before, and select the value of $kT_e$\footnote{For type-A and type-B QPOs, acceptable fits are obtained by fixing $kT_e \sim 10-20$ keV in SIMS.} with minimal change in $\chi^2_{\rm red}$. The \texttt{cov\_frac} is computed respectively as $\sim 0.8$ in LHS, $\sim 0.4-0.5$ in SIMS and $\lesssim 0.02$ in HSS for all sources under considerations. However, for Swift J1727.8$-$1613, we adopt a model combination with an additional cutoff power-law component to compensate for high energy excess following \cite{Yang-etal-2024} as \texttt{tbabs$\times$(thcomp$\otimes$diskbb+ga+cutoffpl)$\times$constant}. We calculate the bolometric luminosity, and estimate the normalized Comptonized flux ($F_{\rm nth}$, ratio between the Comptonized flux and the total flux) and normalized disc flux (ratio between the disc flux and the total flux) for the energy range of $1-100$ keV.

We put efforts to estimate the spin of nine black hole sources using continuum fitting. We employ relativistic accretion disc model \texttt{kerrbb2} \cite[]{McClintock-etal-2006} and calculate the black hole spin considering the hardening factor lies between $1.4-2$ \cite[]{Davis-etal-2005,Davis-El-Abd-2019}. With this, we adopt a model combination as \texttt{tbabs$\times$(simpl$\otimes$kerrbb2)$\times$constant}, where \texttt{simpl} is used for fitting the Comptonization component of the energy spectrum \cite[]{Steiner-etal-2009}. Observations in the HSS are considered for modeling if the scattering fraction, $\text{f}_{\text{scat}}$, is less than or equal to 25\% \cite[]{Steiner-etal-2011}. However, this modeling does not apply to Swift J1753.5$-$0127, Swift J1727.8$-$1613, MAXI J$1535-571$ and Swift J$1658.2-4242$ as the conditions required were not satisfied in any of their observations.

\begin{figure*}
    \begin{center}
    \includegraphics[width=\textwidth]{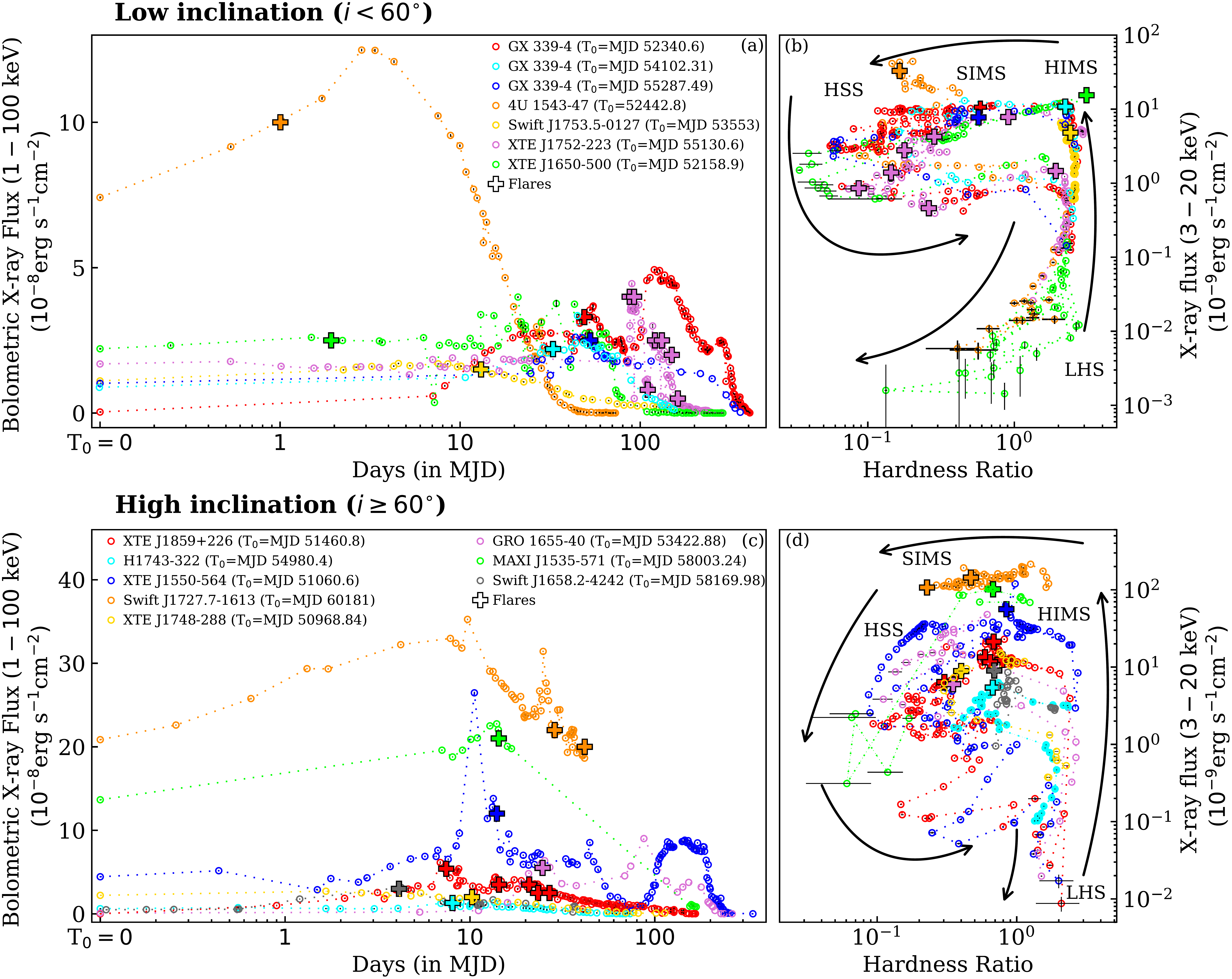}  
    \end{center}
    \caption{Evolution of (a) bolometric ($1-100$ keV) X-ray flux in low inclination sources such as GX $339-4$ (outbursts in $2002$, $2006$ and $2010$ represented by red, cyan and blue symbols, respectively), 4U $1543-47$ (orange), Swift J$1753.5-0127$ (yellow), XTE J$1752-223$ (pink) and XTE J$1650-500$ (light green) and (c) in high inclination sources such as XTE J$1859+226$ (red), H$1743-322$ (2009 outburst) (cyan), XTE J$1550-564$ (blue), Swift J$1727.8-1613$ (orange, using HXMT data), XTE J$1748-288$ (yellow), GRO J$1655-40$ (pink), MAXI J$1535-571$ (light green using HXMT data) and Swift J$1658.2-4242$ (grey, using AstroSat and HXMT data). The start date of the outbursts is denoted by T$_0$. The hardness intensity diagram (HID) for both low and high-inclination sources are illustrated in (b) and (d), respectively. The radio flares are marked using the ‘plus’ symbol in the outburst profile and HID. See the text for details.}
    \label{fig:fig01}
\end{figure*}

\section{Results}
\label{sec:section4}

\subsection{Outburst Profile and Hardness Intensity Diagram}

In Fig. \ref{fig:fig01}, we present the outburst profiles of thirteen BH-XRBs under consideration, where bolometric ($1-100$ keV) X-ray fluxes are plotted as function of day (in MJD) starting from the beginning of the outburst (T$_{0}$). Open symbols connected with dotted lines denote the outburst profile for GX $339-4$ (red, cyan and blue for 2002, 2007 and 2010 outbursts), 4U 1543$-$47 (orange; 2002 outburst), Swift J1753.5$-$0127 (yellow; 2005 outburst), XTE J1752$-$223 (pink; 2009 outburst), XTE J1650$-$500 (light green; 2001 outburst), XTE J1859$+$226 (red; 1999 outburst), H1743$-$322 (cyan; 2009 outburst), XTE J1550$-$564 (blue; 1998 outburst), Swift J1727.8$-$1613 (orange; 2023 outburst), XTE J$1748$-$288$ (yellow; 1998 outburst), GRO J$1655-40$ (pink; 2005 outburst), MAXI J$1535-571$ (light green; 2017 outburst) and Swift J$1658.2-4242$ (grey; 2018 outburst) sources, respectively. `Plus' symbol indicates the time of radio flares observed during an outburst. At the inset, we depict HID for all the sources using same point styles and colors as used for displaying outburst profiles.

\begin{figure}
    \begin{center}
    \includegraphics[width=\columnwidth]{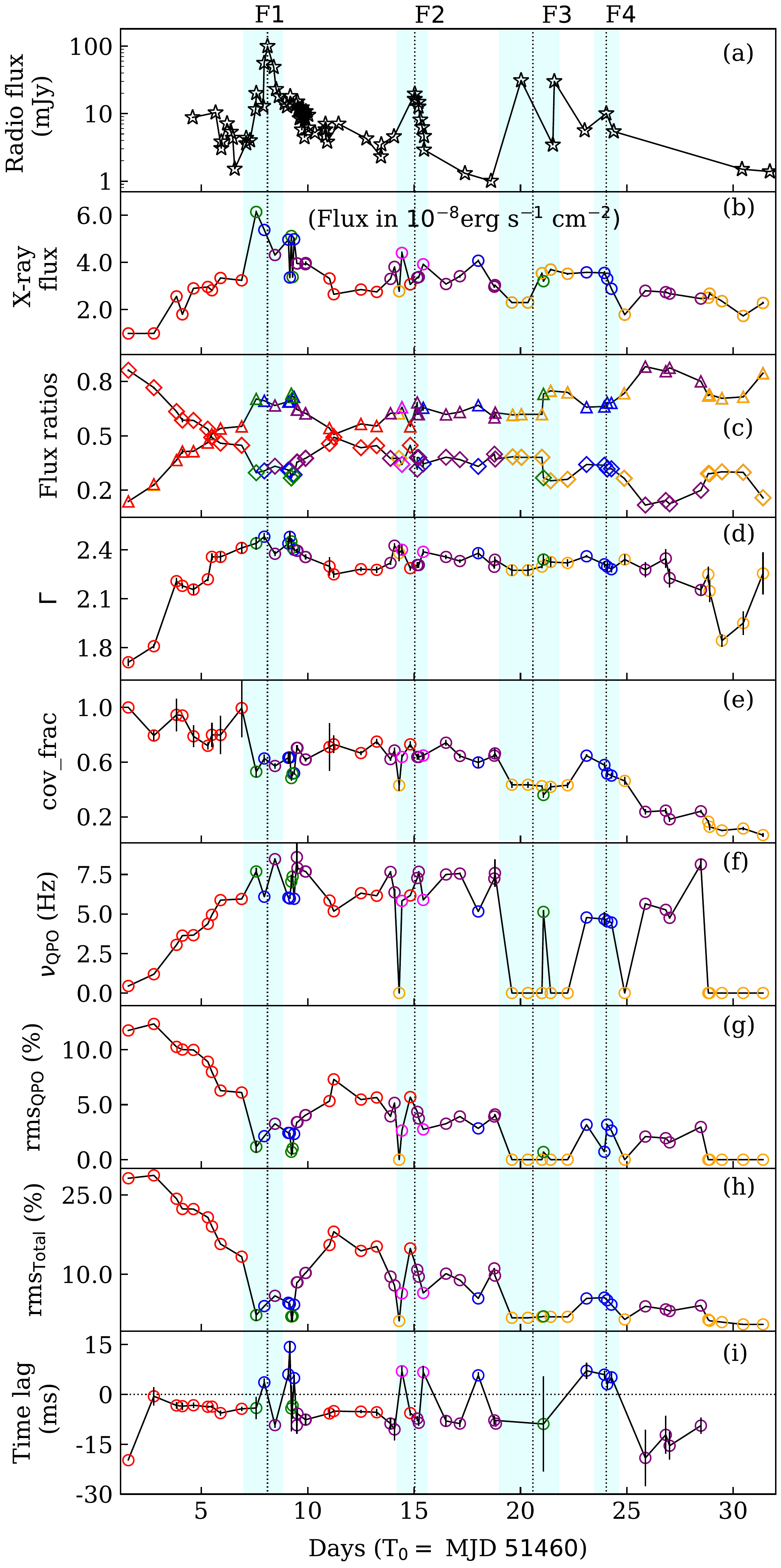}  
    \end{center}
    \caption{Evolution of (a) Radio flux in $1.4-1.66$ GHz (asterisk), (b) bolometric X-ray flux ($1-100$ keV), (c) Flux ratios (normalized disc flux (triangle) and normalized Comptonized flux (diamond)), (d) photon index ($\Gamma$), (e) covering fraction (cov\_frac), (f) centroid frequency of QPO ($\nu_{\text{QPO}}$), (g) QPO rms (rms$_{\text{QPO}}$), (h) total rms (rms$_{\text{Total}}$), and (i) time lag at QPO frequency between energy range $6-15$ keV and $2-6$ keV  of XTE J1859$+$226 during its 1999 outburst. Data points plotted with various colors represent different types of QPO (green: type-A QPO, blue: type-B QPO, magenta: type-B cathedral, red: type-C QPO and purple: type-C* QPO, orange: absence of QPO). The vertical dotted lines represent the peak of radio flares. Shaded regions represent the flaring zone consisting of rising phase, peak value and declining phase of radio flux in a particular flare. See the text for details.}
    \label{fig:fig02} 
\end{figure}

During $1999$ outburst, XTE J1859$+$226 traced all the canonical spectral states, such as LHS, HIMS, SIMS and HSS, in its hysteresis loop depicted by open circles (in red) in HID of Fig. \ref{fig:fig01}. The source exhibited its peak X-ray flux ($\sim 6 \times 10^{-8}~\text{erg} ~\text{s}^{-1}\text{cm}^{-2}$) along with multiple radio flares \cite[]{Brocksopp-etal-2002, Fender-etal-2009, Radhika-Nandi-2014} during SIMS (HR $\lesssim 0.7$) as indicated by `plus' symbol in Fig. \ref{fig:fig01} and dotted vertical lines (marked as F1, F2, F3 and F4) in Fig. \ref{fig:fig02}. We observe a slight increase in X-ray flux preceding each radio flare. Similar to XTE J1859$+$226, GX 339$-$4 during its 2002, 2007 and 2010 outburst traced all canonical spectral states. It exhibited one X-ray peak in 2007 and 2010 outbursts with X-ray fluxes $\sim 3.3 \times 10^{-8}$ erg s$^{-1}$ cm$^{-2}$ and $\sim 2.6 \times 10^{-8}$ erg s$^{-1}$ cm$^{-2}$, respectively. During its 2002 outburst, it exhibited two peaks in X-rays (during SIMS with flux ($\sim 3.5\times10^{-8}~\text{erg}~ \text{s}^{-1}\text{cm}^{-2}$ and HSS with flux $\sim5.0\times10^{-8}\text{erg} \text{s}^{-1}\text{cm}^{-2}$), however radio flare is observed $\sim 3$ days before the X-ray peak in SIMS (HR $\sim 0.6$; Fig. \ref{fig:fig01} and Fig. \ref{fig:fig04}). The source 4U 1543$-$47 was observed in SIMS while transiting from HIMS, showing a single X-ray peak ($\sim1.7\times10^{-7}~\text{erg} ~\text{s}^{-1}\text{cm}^{-2}$) during SIMS, with a radio flare observed approximately 2 days before the peak X-ray flux (HR $\ge 0.1$; Fig. \ref{fig:fig01} and Fig. \ref{fig:fig06}). H1743$-$322 is seen to transit from HIMS to SIMS with a peak X-ray flux ($\sim1.6\times10^{-8}~\text{erg} ~\text{s}^{-1}\text{cm}^{-2}$) during SIMS. As before, we find that the radio flare occurs $\sim 1$ day after the peak X-ray flux at HR $\sim 0.6$ (see Fig. \ref{fig:fig01}). Similar to GX 339$-$4, XTE J1550$-$564 also evolved from HIMS to SIMS with multiple X-ray peaks ($\sim2.8\times10^{-7}~\text{erg} ~\text{s}^{-1}\text{cm}^{-2}$ during SIMS and $\sim1.3\times10^{-7}~\text{erg} ~\text{s}^{-1}\text{cm}^{-2}$ during HSS). The radio flare is observed approximately $1$ day after the peak X-ray flux in the SIMS phase, when the hardness ratio (HR) is $\gtrsim0.8$ (see Fig. \ref{fig:fig01}).

The source XTE J1752$-$223 exhibited an outburst in 2009, but due to Sun constraints, no observations were available from MJD $55155.09-55215.90$. During its outburst, the source reached a peak X-ray flux of $\sim 5 \times 10^{-8}$ erg s$^{-1}$ cm$^{-2}$, accompanied by multiple radio flares \cite[]{Brocksopp-etal-2013} in the SIMS, HSS, and during the transition from SIMS to HIMS in the decay phase. XTE J1752$-$223 displayed all canonical spectral states, with marginal rise in X-ray flux near the radio flares. XTE J1650$-$500 was also observed in all canonical spectral states with the peak X-ray flux of $4 \times 10^{-8}$ erg s$^{-1}$ cm$^{-2}$ occurring during SIMS. The peak radio flux of $\sim 5.28$ mJy at $4.8$ GHz was observed during HIMS, though radio observations were absent during the transition from HIMS to SIMS, potentially missing the flare event \cite[]{Fender-etal-2009}. Swift J1753.5$-$0127 underwent a failed outburst and evolve from HIMS to LHS in the HID. Similar to other canonical outbursting sources, the peak radio flux of $2.48$ mJy at $8.4$ GHz coincided with the X-ray peak flux of $2 \times 10^{-8}$ erg s$^{-1}$ cm$^{-2}$. Recently, Swift J1727.8$-$1613 exhibited the highest X-ray flux among the sources, reaching a peak of $\sim 3.5 \times 10^{-7}$ erg s$^{-1}$ cm$^{-2}$. Multiple radio flares ($\ge 100$ mJy at $334$ MHz) were observed as the source transits to SIMS. For GRO~J1655$-$40, the source exhibits all four canonical spectral states, with two distinct peaks in X-ray flux: $\sim 6.4 \times 10^{-8}$~erg~s$^{-1}$~cm$^{-2}$ during SIMS and $\sim 9.5 \times 10^{-8}$~erg~s$^{-1}$~cm$^{-2}$ during HSS. XTE~J1748$-$288 also makes transition through all four spectral states, reaching a peak X-ray flux of $\sim 2.5 \times 10^{-8}$~erg~s$^{-1}$~cm$^{-2}$. MAXI~J1535$-$571 is observed in LHS, HIMS, SIMS, and HSS with a maximum X-ray flux of $\sim 22 \times 10^{-8}$~erg~s$^{-1}$~cm$^{-2}$ during SIMS. Swift~J1658.2$-$4242, on the other hand, is observed during the HIMS and SIMS phases, exhibiting a peak X-ray flux of $\sim 3.1 \times 10^{-8}$~erg~s$^{-1}$~cm$^{-2}$.

It is worth mentioning that XTE J1859$+$226, H1743$-$322, XTE J1550$-$564, Swift J1727.8$-$1613, XTE J$1748-288$, GRO J$1655-40$, MAXI J$1535-571$ and Swift J$1658.2-4242$ are high-inclination sources ($i \ge 60^\circ$) \cite[]{Hynes-etal-2002, Zurita-etal-2002, Jonker-etal-2004, Shafee-etal-2006, Jonker-etal-2010, Orosz-etal-2011,  Steiner-etal-2012, Corral-Santana-etal-2013, Eijnden-etal-2017, Miller-etal-2018, Xu-etal-2018, Rizo-etal-2022, Wood-etal-2024,Yu-etal-2024, Abdulghani-etal-2024}, and these sources typically show radio flares about one day after the X-ray peak in SIMS. In contrast, GX 339$-$4, 4U 1543$-$47, XTE J1752$-$223, XTE J1650$-$500, and Swift J1753.5$-$0127 are low-inclination sources with $i \sim 35^\circ - 50^\circ$.

\subsection{Evolution of Spectro-temporal Properties During Jet Ejection}
\label{sec:section4.2}

In this section, we study the evolution of spectro-temporal	parameters during the entire outburst of all the sources under consideration (XTE J$1859+226$, GX $339-4$ (2002, 2007 and 2010 outbursts), 4U $1543-47$, H$1743-322$ (2009 outburst), XTE J$1550-564$, XTE J$1752-223$, XTE J$1650-500$, Swift J$1753.5-0127$, XTE J$1748-288$, GRO J$1655-40$, Swift J$1727.8-1613$, MAXI J$1535-571$ and Swift J$1658.2-4242$.) In Fig. \ref{fig:fig01}, we denote the peak radio flux with `plus' symbol which indicates the presence of the radio ejection in SIMS (as shown in the inset).

In Fig. \ref{fig:fig02}, we present the evolution of spectro-temporal parameters of XTE J$1859$+$226$ during $1999$ outburst. We depict the evolution of Radio flux, bolometric X-ray flux, Flux ratios (normalized disc and Comptonized fluxes), photon index ($\Gamma$), covering fraction (cov\_frac), centroid frequency ($\nu_{\rm QPO}$), QPO rms (rms$_{\rm QPO}\%$), total rms (rms$_{\rm Total}\%$) and time lag at QPO frequency with time since the triggering of the outburst. XTE J1859$+$226 exhibits four radio flares (marked as F1, F2, F3 and F4 at the top of Fig. \ref{fig:fig02}) along with X-ray peaks during SIMS (see Fig. \ref{fig:fig01}). 

\begin{figure}
    \begin{center}
    \includegraphics[width=\columnwidth]{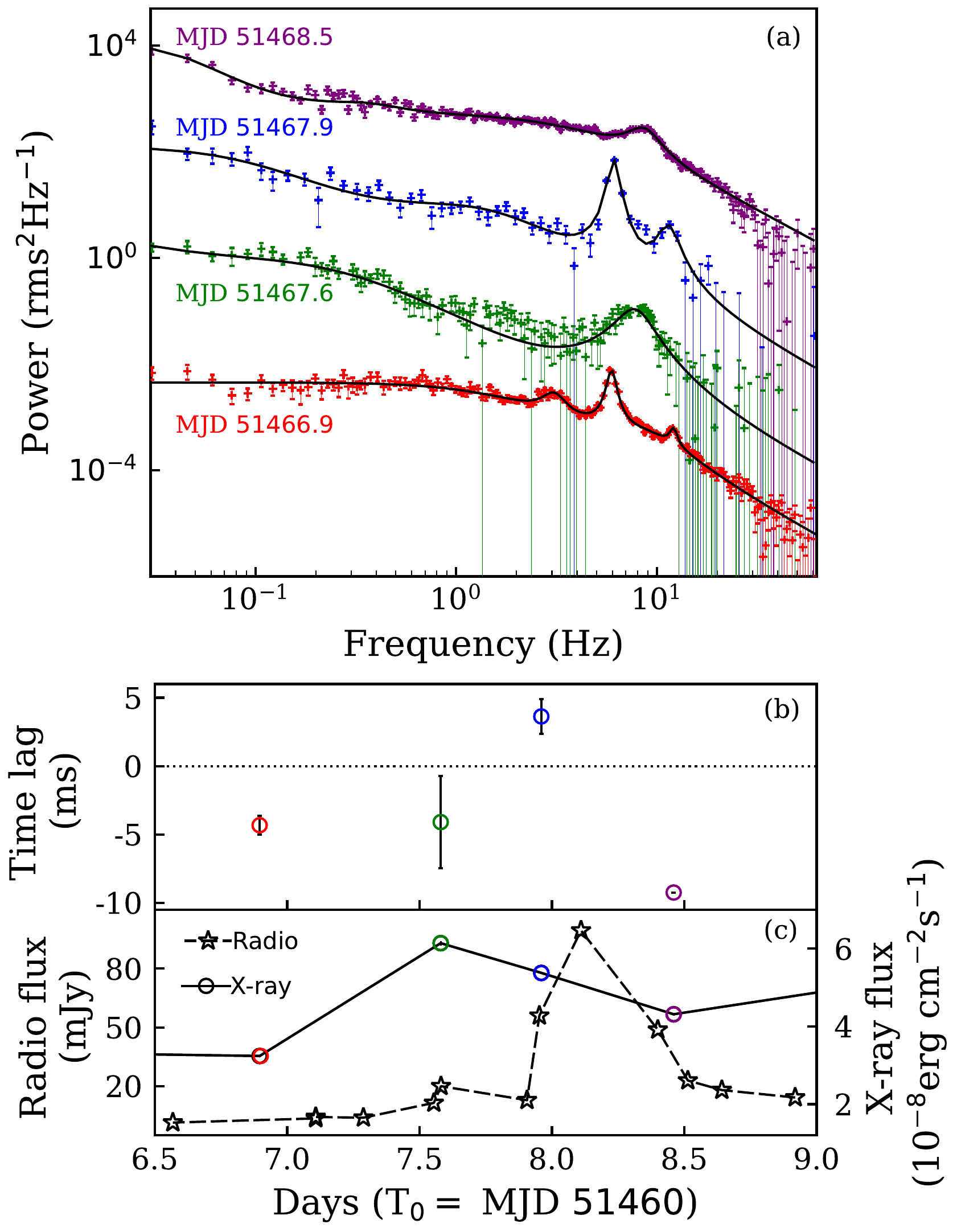}  
    \end{center}
    \caption{Evolution of (a) PDS (scaled for better clarity), (b) time lag (open circle) corresponding to different types of QPOs, and (c) bolometric X-ray flux (open circle) and Radio flux in $1.4-1.66$ GHz (asterisk) near flare F1 (MJD 51467.9) of XTE J1859$+$226. Different colors denote various types of QPO (green: type-A, blue: type-B, red: type-C and purple: type-C*). See the text for details.}
    \label{fig:fig03}
\end{figure}

During the rising phase of the outburst, XTE J1859$+$226 exhibits type-C QPOs (evolves from $0.45-6.20$ Hz) with weak variation of radio emission ($\sim 10$ mJy). When the source transits from HIMS to SIMS, a type-A QPO appears along with a peak X-ray flux ($\sim6.14\times10^{-8}$ erg cm$^{-2}$ s$^{-1}$) and enhanced radio flux {$100$ mJy, indicating radio ejection marked as F1 (see also Fig. \ref{fig:fig03}). It’s worth noting that type-B/B-cathedral QPOs appear about $3-10$ hours before the radio flares F1, F2, F3, F4, which are marked by a positive time lag of around $5-8$ ms. Generally, observations show type-A QPOs, weak type-C* QPOs, or the absence of QPOs prior to the appearance of type-B QPOs, with rms$_{\rm QPO}\%$ around $1.0\%$, compared to $2.5\%$ for type-B QPOs. In addition, type-A and type-C* QPOs are associated with a negative time lag of $\sim 5$ ms. Throughout the flaring periods, the disc flux ($\ge 0.5$) is observed to dominate over the Comptonized X-ray flux. Next, we focus on the evolution of PDS during the radio flare, specifically for flare F1, and examine how it correlates with variations in radio flux, X-ray flux, and time lag (see Fig. \ref{fig:fig03}). As the source transits from HIMS to SIMS, it exhibits a type-C QPO on MJD 51466.9, followed by a type-A QPO on MJD $51467.6$, type-B on MJD 51467.9 and type-C* on MJD $51468.5$, respectively (see panel a). The time lag in these four observations flips from negative to positive (see panel (b)), coinciding with the onset of a radio flare. As QPO type alters from type-A to type-B, both X-ray flux (depicted by open circles) and radio flux (marked by asterisks) increases (see panel c). These results possibly suggest a change in the accretion dynamics in the vicinity of the source under consideration. After $\sim 10$ hrs, the nature of the QPO is again altered to type-C* yielding a negative time lag. Similar time lags associated with type-B QPOs are also observed for F2, F3 and F4 radio flares as well. Although the type-A QPO is observed before the radio ejection, type-B QPOs are observed with positive time lag for all four radio flares.

We investigate the remaining seven high-inclination sources, namely H1743$-$322 \cite[]{Steiner-etal-2012, Molla-etal-2017}, XTE J1550$-$564 \cite[]{Jonker-etal-2004, Jonker-etal-2010, Orosz-etal-2011}, Swift J1727.8$-$1613 \cite[]{Wood-etal-2024,Yu-etal-2024}, GRO J$1655-40$ \cite[]{Markwardt-etal-2005}, XTE J$1748-288$ \cite[]{Brocksopp-etal-2007}, MAXI J$1535-571$ \cite[]{Negoro-etal-2017} and Swift J$1658.2-4242$ \cite[]{Grebenev-etal-2018}. During the initial phase of $2009$ outburst of H1743$-$322, type-C QPOs are observed in HIMS with $\nu_{\rm QPO}$ increasing from $0.9-3.6$ Hz. Subsequently, type-A QPO appears as the sources transits from HIMS to SIMS. The appearance of type-A QPO ($\nu_{\rm QPO} \sim 3.4$ Hz, rms$_{\rm QPO}\% \sim 2\%$ and rms$_{\rm Total}\% \sim 6.1 \%$) during state transition coincides with the observation exhibiting the peak X-ray flux in the entire outburst. After $\sim 1$ day, the radio flare ($\sim 7$ mJy at $8.46$ GHz) is observed. The subsequent observation in X-ray band does not exhibit QPO in the PDS followed by type-C* QPO ($\nu_{\rm QPO} \sim 6.8$ Hz, rms$_{\rm QPO}\% \sim 2.5\%$ and rms$_{\rm Total}\% \sim 6.6\%$) and type-B QPO ($\nu_{\rm QPO} \sim 3.7$ Hz, rms$_{\rm QPO}\% \sim 3.5\%$ and rms$_{\rm Total}\% \sim 7.2\%$). Type-C, type-C* and type-A QPOs display negative time lags ($\sim 0.3-2$ ms, $\sim 3$ ms and $\sim 10$ ms), whereas type-B QPOs demonstrate positive time lag ($\sim 5 \pm 2.8$ ms) during 2009 outburst. The appearance of type-A QPO during the state-transition corresponding to the peak X-ray flux just before the appearance of radio flare marks the onset of ejection phenomena.

\begin{figure}
    \begin{center}
    \includegraphics[width=\columnwidth]{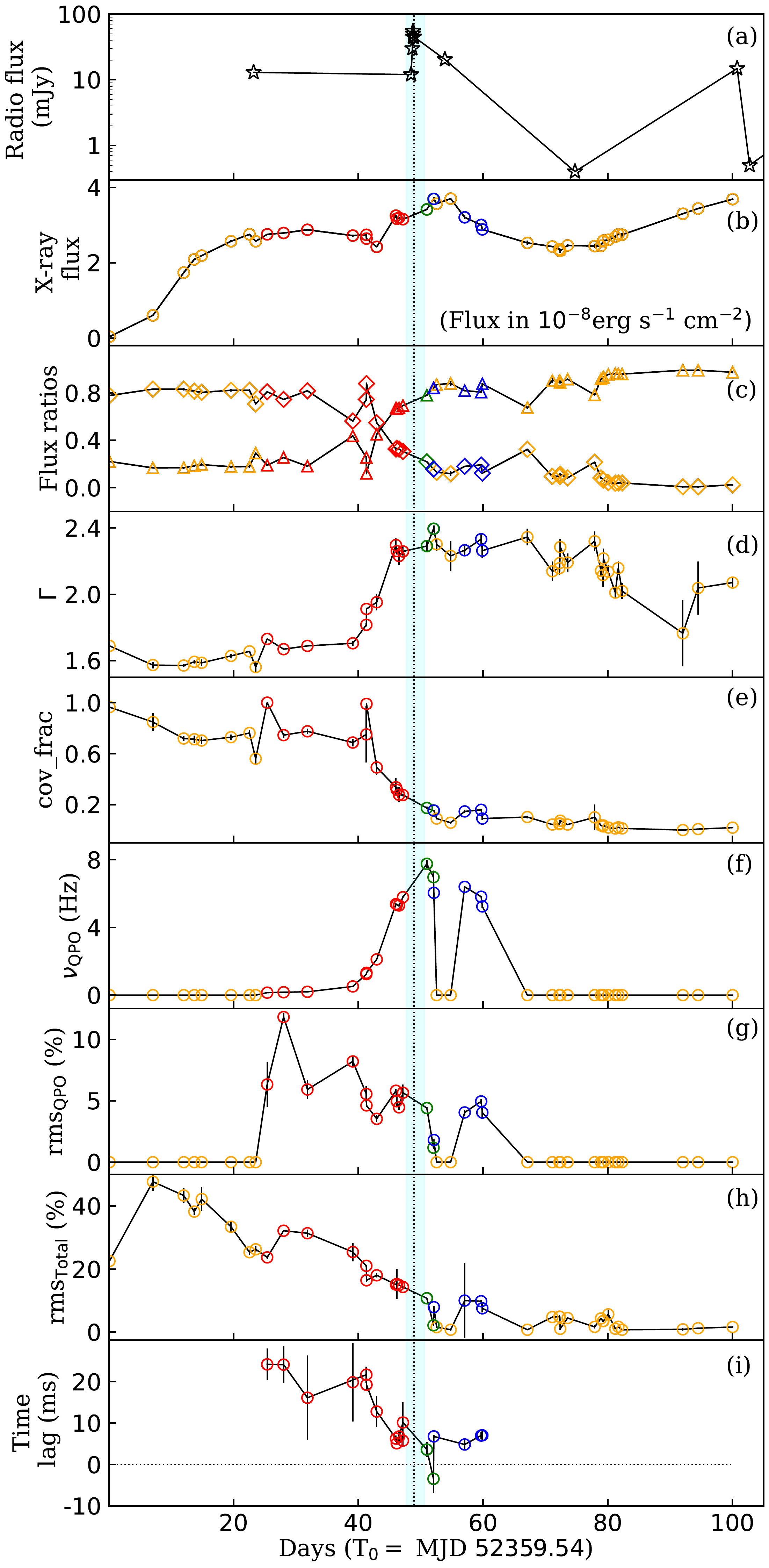}  
    \end{center}
    \caption{Same as Fig. \ref{fig:fig02}, but for GX 339$-$4 during its $2002-2003$ outburst. See the text for details.}
\label{fig:fig04}
\end{figure}

During $1998$ outburst, XTE J1550$-$564 underwent spectral state transition from HIMS to SIMS with the detection of radio flare ($234$ mJy at $8.6$ GHz) and appearance of type-A QPO ($\nu_{\rm QPO} \sim 12$ Hz and rms$_{\rm QPO}\% \sim 0.5 \%$ and rms$_{\rm Total}\% \sim 2\%$). Type-C QPO again appears in the PDS after the radio flare, with $\nu_{\rm QPO}$ decreasing from $\sim 5.4$ Hz to $\sim 2.6$ Hz, then rising again from $\sim 2.7$ Hz to $\sim 5.2$ Hz. Interestingly, type-A and type-B QPOs re-appear in the PDS just before the source transits from SIMS to HSS, with $\nu_{\rm QPO}$ values of $\sim 6.9$ Hz and $\sim 4.5$ Hz, respectively. The time lag associated with type-C QPOs varies between positive and negative values, while the type-A QPOs show a negative lag of $\sim 5$ ms, and type-B QPOs exhibit time lag as $-2.3 \pm 1.6$ ms (MJD $51106.95$), $1.7 \pm 0.5$ ms (MJD $51108.07$) and $ -4.0 \pm 2.6 $ ms (MJD $51110.26$), respectively.

Swift J1727.8$-$1613 shows type-C QPOs with monotonically increasing frequency as $0.37-8.74$ Hz before F1 radio flare \cite[]{Nandi-etal-2024} during $2023$ outburst. The appearance of the type-B ($\nu \sim 6.97$ Hz, rms$_{\rm QPO}\% \sim 4.91\%$) QPO on MJD $60206.13$, prior to F1, suggests that this observation marks the transition from HIMS to SIMS as indicated by \cite{Nandi-etal-2024} based on the HR value $\sim 0.19-0.02$ using \textit{MAXI/GSC} observations. After F1, $\nu_{\rm QPO}$ increases till $8.85$ Hz and finally disappears before F2 (see Fig. \ref{fig:fig06}). We observe both positive and negative time lags associated to type-C QPOs with high normalized Comptonized flux ($F_{\rm nth}$) (see Fig. \ref{fig:fig07}).

Next, we examine the spectro-temporal characteristics of low-inclination source, namely GX 339$-$4. We find that GX 339$-$4 exhibits only one radio flare during the entire $2002-2003$ outburst. In Fig. \ref{fig:fig04}, we present the evolution of the timing as well as spectral parameters obtained during the outburst. As the outburst progresses, the frequency of type-C QPOs increases with the enhancement of the X-ray flux with peak value $\sim 3.5 \times 10^{-8}~\text{erg~ s}^{-1}\text{cm}^{-2}$. During  the transition from HIMS to SIMS, radio flare is observed with peak flux ($55$ mJy) and type-A QPO appears followed by type-B QPO. It is worth mentioning that the time lag decreases from $\sim 21$ ms to $\sim-3.5$ ms, where negative lag is observed corresponding to one of the type-A QPOs. Furthermore, we notice that the normalized Comptonized flux and disc flux alter before the radio flare. In Fig. \ref{fig:fig05}, we depict the evolution of the PDS along with the X-ray and radio fluxes during the radio flare (MJD $52410.5$ to MJD $52412.1$) of GX 339$-$4. We observe that type-A and type-B QPO emerge $\sim 2$ days after the radio flare. We infer that this delay possibly arises due to a lack of X-ray monitoring. Furthermore, we also analyzed 2007 and 2010 outbursts of GX 339$-$4 and the obtained results are presented in Figs. \ref{fig:fig06}-\ref{fig:fig08}.

\begin{figure}
    \begin{center}
    \includegraphics[width=\columnwidth]{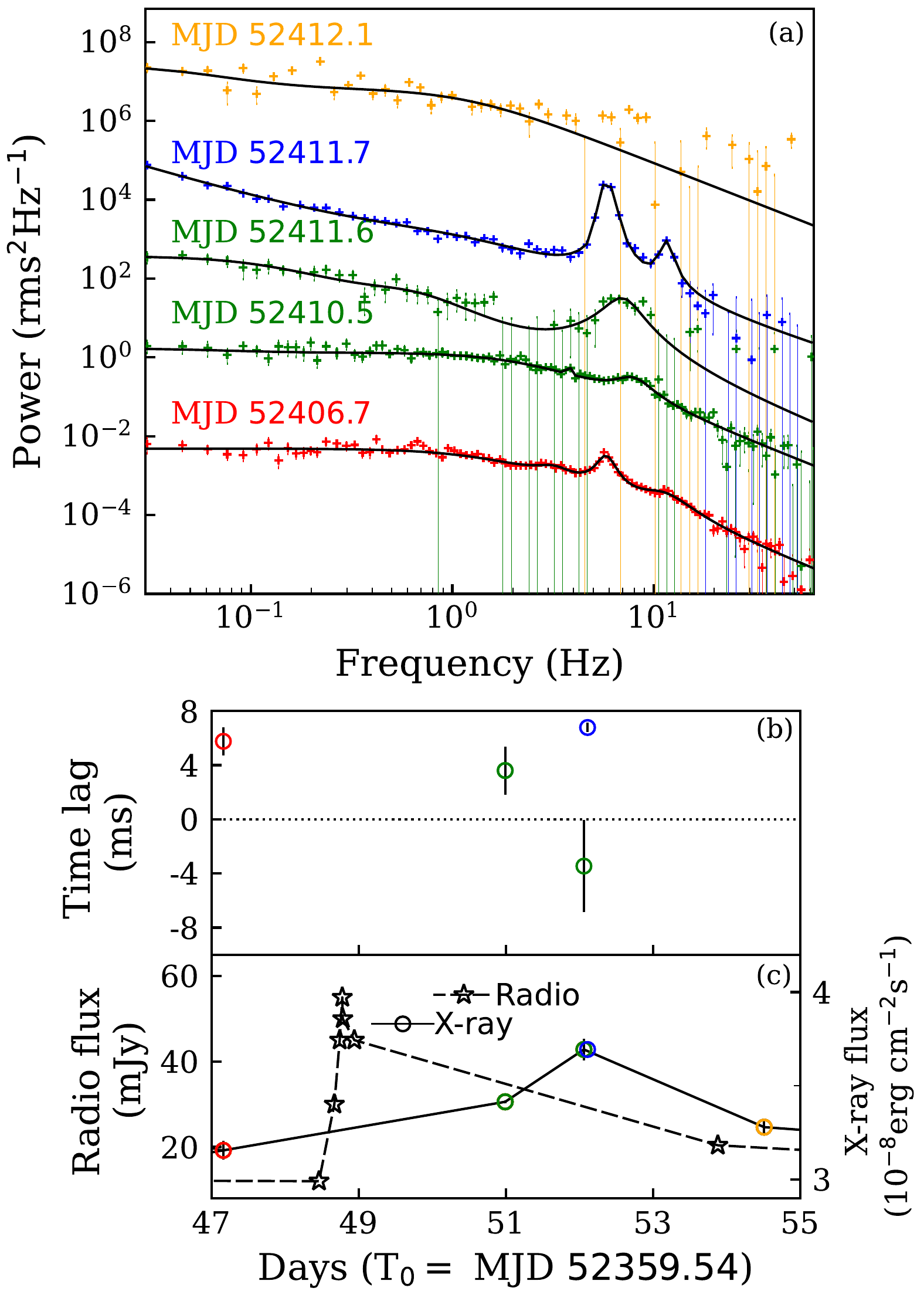}
    \end{center}
    \caption{Evolution of (a) PDS (scaled for better clarity), (b) time lag (open circle) corresponding to different types of QPOs, and (c) bolometric X-ray flux (open circle) and Radio flux at $4.8$ GHz (asterisk) near flares (MJD $52408.31$) of GX 339$-$4. Different colors denote various types of QPO (green: type-A, blue: type-B, red: type-C and orange: absence of QPO. See the text for details.}
    \label{fig:fig05}
\end{figure}

4U 1543$-$47 is another low-inclination source with $i \sim 20^{\circ}-40^{\circ}$ \cite[]{Orosz-etal-1998, Chen-Wang-2024}. During $2002$ outburst, the source transits from HIMS to SIMS exhibiting peak X-ray flux ($\sim 12.5 \times 10^{-8}$ erg s$^{-1}$ cm$^{-2}$) accompanied by a radio flare ($\sim 22$ mJy at $1026.75$ MHz). After the radio flare, QPO disappears from the PDS. A subsequent weak radio flare ($\le 0.3$ mJy) was detected \cite[]{Russell-etal-2020}, and type-A and type-B QPOs were observed near the time of the radio flare (see Fig. \ref{fig:fig06}). Type-B QPO displays positive time lags ($\sim 2.3-6.2$ ms) and type-A QPO is observed with negative time lags ($\sim 1.7-5.5$ ms). Note that a type-C QPO of $\nu_{\rm QPO} \sim 9$ Hz is observed towards the end of the outburst.

XTE J1650$-$500 exhibits a canonical outburst, with the peak radio flux of approximately $\sim 5.3$ mJy at $4.8$ GHz and a spectral index of $\sim -0.27$ observed during the hard intermediate state (HIMS). Type-C QPOs are observed to evolve with $\nu_{QPO} \sim 0.97-6.88$ Hz. Near the HIMS to SIMS transition, type-B QPOs ($\nu_{\rm QPO} \sim 16-17$ Hz, rms$_{\rm QPO} \sim 3\%$) appear and then disappear in the PDS, with no radio observations available near the transition (see Fig. \ref{fig:fig06}). Both type-C and type-B QPOs exhibit a positive lag.  

Swift J1753.5$-$0127 displays a failed outburst, remaining in LHS and HIMS, with its HID mimicking the decay phase of a canonical outburst (see Fig. \ref{fig:fig01}). At the beginning of the outburst, the source is in HIMS, as indicated by its high bolometric flux ($\ge 1.0 \times 10^{-8}$ erg cm$^{-2}$ s$^{-1}$) and HR ($\le 2.4$). The peak radio flux ($\ge 1.5$ mJy at $8.4$ GHz with a spectral index of $\sim -0.08$) is observed near the peak X-ray flux. The source transits into LHS during the decay phase of the outburst. Type-C QPOs are observed with positive time lags throughout the outburst and $\nu_{\rm QPO}$ is seen to increase from $0.60$ Hz to $\sim 0.85$ Hz at the beginning and then decreases during the decay phase.

XTE~J1748$-$288 underwent a canonical outburst during its 1998 outburst. Observations commenced while the source was in HIMS, during which it exhibited a peak bolometric X-ray flux of $\sim 2.7 \times 10^{-8}$~erg~s$^{-1}$~cm$^{-2}$. At the onset of the outburst, the PDS revealed the presence of type-C QPO with centroid frequencies in the range $\nu \sim 17-31$~Hz and fractional rms amplitudes of $\sim 1.7\% - 4.3\%$. The Q factor associated with these QPOs spans $\sim 3-12$. Note that the QPOs vanish from the PDS prior to the detection of a radio flare.

During its 2005 outburst, GRO~J1655$-$40 exhibited all four canonical spectral states, reaching a peak bolometric X-ray flux of $\sim 6.4 \times 10^{-8}$~erg~s$^{-1}$~cm$^{-2}$ during the transition from HIMS to SIMS. In the early phase of the outburst, type-C QPOs were detected during the HIMS, with centroid frequencies ($\nu_{\rm QPO}$) increasing from $\sim 0.11$~Hz to $2.3$~Hz and corresponding fractional rms amplitudes rising from $\sim 4\%$ to $14\%$. As the source was transited from HIMS to SIMS, the type-C QPOs disappeared, and a type-B QPO emerged with $\nu_{\rm QPO} \sim 6.4$~Hz and rms amplitude of $\sim 4\%$ during the SIMS.

We have analyzed 2018 outburst of MAXI J1535$-$571 using \textit{Insight/HXMT} data. The source exhibits type-C QPOs during HIMS where $\nu_{\rm QPO}$ increases from $\sim 2.1$ Hz to $8.9$ Hz and rms$_{\rm QPO}\%$ decreases from $5.8\%$ to $3.2\%$.  Type-A QPO is observed in the SIMS at $\nu_{\rm QPO}\sim 6.3$ Hz and with rms$_{\rm QPO}\sim 0.99\%$ before the radio flare.

Swift~J1658.2$-$4242 exhibits type-C QPOs with centroid frequencies ranging from $\sim 1.5$ to $6.6$~Hz during the rising phase of the outburst. Near the peak of the outburst, a type-A QPO is detected at $\nu_{\rm QPO} \sim 6-7$~Hz with a fractional rms amplitude of $\sim 2-5\%$. Additionally, a type-B QPO is observed at $\nu_{\rm QPO} \sim 4$~Hz, exhibiting a higher fractional rms amplitude of $\sim 6-8\%$. 

\begin{figure}
    \begin{center}
    \includegraphics[width=\columnwidth]{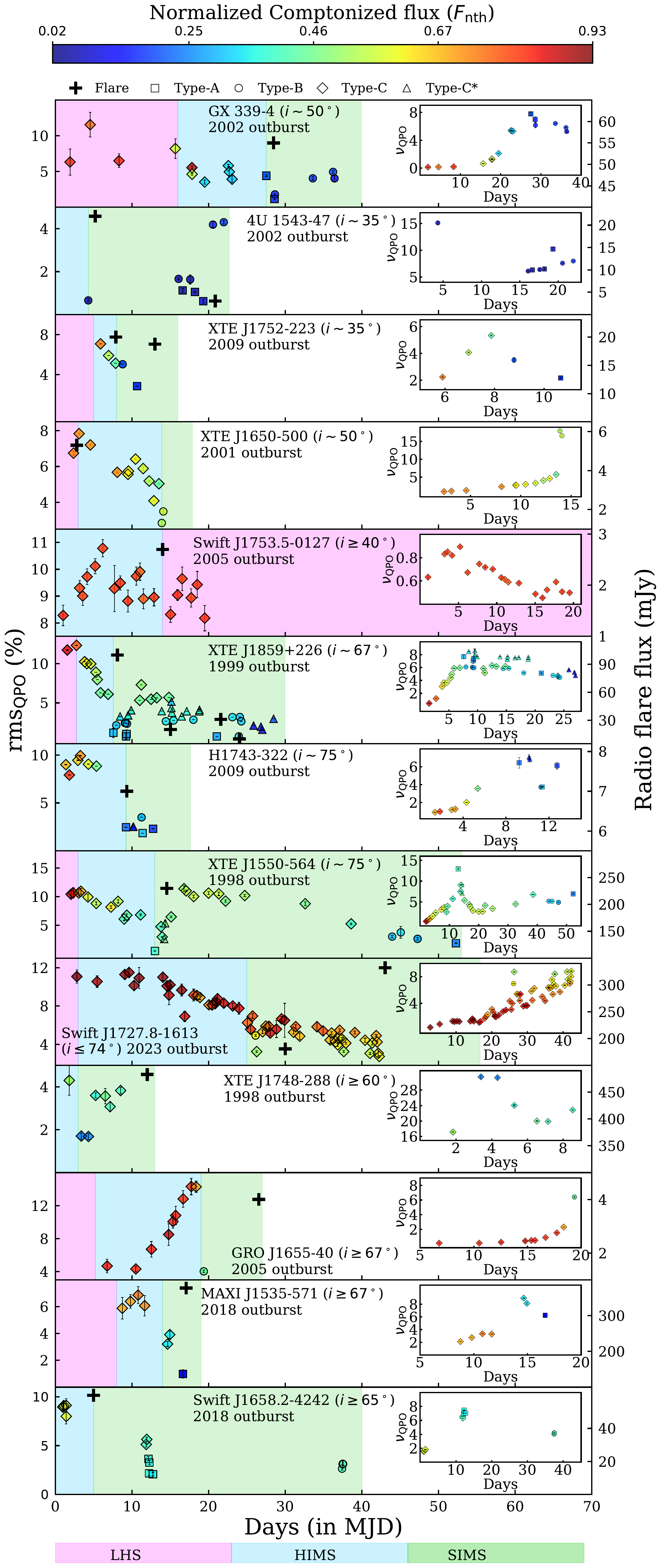}
    \end{center}
    \caption{Evolution of QPO rms (rms$_{\text{QPO}}\%$) with time in GX $339-4$ ($i\sim50^\circ$), 4U $1543-47$ ($i\sim36^\circ$), XTE J$1752-223$ ($i\sim35^\circ$), XTE J$1650-500$ ($i\ge47^\circ$), Swift J$1753.5-0127$ ($i\ge40^\circ$), XTE J$1859+226$ ($i\sim65^\circ$), H$1743-322$ ($i\sim75^\circ$), XTE J$1550-564$ ($i\sim74^\circ$), Swift J$1727.8-1613$ ($i\le74^{\circ}$), XTE J$1748-288$ ($i\ge60^{\circ}$), GRO J$1655-40$ ($i\ge67^{\circ}$), MAXI J$1535-571$ ($i\ge67^{\circ}$) and Swift J$1658.2-4242$ ($i\ge65^{\circ}$). Colorbar represents the normalized Comptonised flux ($F_{\rm nth}$). Square, circle, diamond and triangle denote different QPO types, and radio flux is indicated using plus symbol. Variation of $\nu_{\rm QPO}$ with days are shown in the inset. The presence of radio flux is indicated using plus symbol. For each sources, the duration of the various spectral states are indicated using different shades. See the text for details.} 
    \label{fig:fig06}
\end{figure}

\subsection{Dependence of Source Inclination on Spectro-Temporal Properties}

In the previous section, we study the evolution of spectro-temporal properties of thirteen outbursting BH sources under consideration. Among them, five are low-inclination systems (GX 339$-$4, 4U 1543$-$47, XTE J1752$-$223, XTE J1650$-$500, Swift J1753.5$-$0127), while the remaining eight are high-inclination systems (XTE J1859$+$226, H1743$-$322, XTE J1550$-$564, Swift J1727.8$-$1613, XTE J$1748-288$, GRO J$1655-564$, MAXI J$1535-571$ and Swift J$1658.2-4242$). For these sources, we examine the correlation of spectro-temporal properties ($i.e.$, rms$_{\rm QPO}\%$, normalized Comptonized flux $F_{\rm nth}$, time lag and energy dependent time lag) during the radio flares including spectral state transitions. Subsequently, we put efforts to decode the imprint of inherent coupling between corona and jets for these sources with varied inclination angles ($\sim 35^\circ - 75^\circ$).

In Fig. \ref{fig:fig06}, we illustrate the evolution of $\rm rms_{QPO}\%$ as well as $\nu_{\rm QPO}$ (see the insets) with time during the rising phase of the outburst, highlighting different spectral states --- LHS, HIMS, and SIMS --- denoted by different shaded regions in the respective panels. The corresponding normalized Comptonized flux ($F_{\rm nth}$) variations for each source are shown using different colors, where $F_{\rm nth}$ lies in the range $\sim 0.02-0.93$ as indicated by the colorbar at the top of the figure. For low-inclination sources, type-A and/or type-B QPOs are generally observed during the radio flares particularly when source transits from HIMS to SIMS, except for Swift J1753.5$-$0127, where the source transits from HIMS to LHS exhibiting only type-C QPOs during the decay phase of the outburst. Similarly, for high-inclination sources, radio flares are observed during the transition from HIMS to SIMS with the clear presence of type-A and/or type-B QPOs. Moreover, absence of QPO is also occasionally observed after the radio flare in SIMS (see Fig. \ref{fig:fig06}). We further notice that type-A and/or type-B QPOs are often observed during the SIMS with relatively lower Comptonized flux values for majority of the sources except for sources, namely XTE J1550$-$564, Swift J1727.8$-$1613, XTE J1859$+$226 and H1743$-$322 (2003 outburst), where type-C QPOs are also detected. We observe that $\rm rms_{QPO}\%$ and $F_{\rm nth}$ are significantly higher for type-C QPOs (usually observed in LHS and HIMS) compared to type-A and type-B QPOs. Near the radio flare, both $\rm rms_{QPO}\%$ and $F_{\rm nth}$ associated with type-A and type-B are decreased. In fact, for all sources, $\rm rms_{QPO}\%$ for type-C QPOs increases with $F_{\rm nth}$ indicating that type-C QPOs possibly originate from the Comptonized corona.

\begin{figure}
    \begin{center}
    \includegraphics[width=\columnwidth]{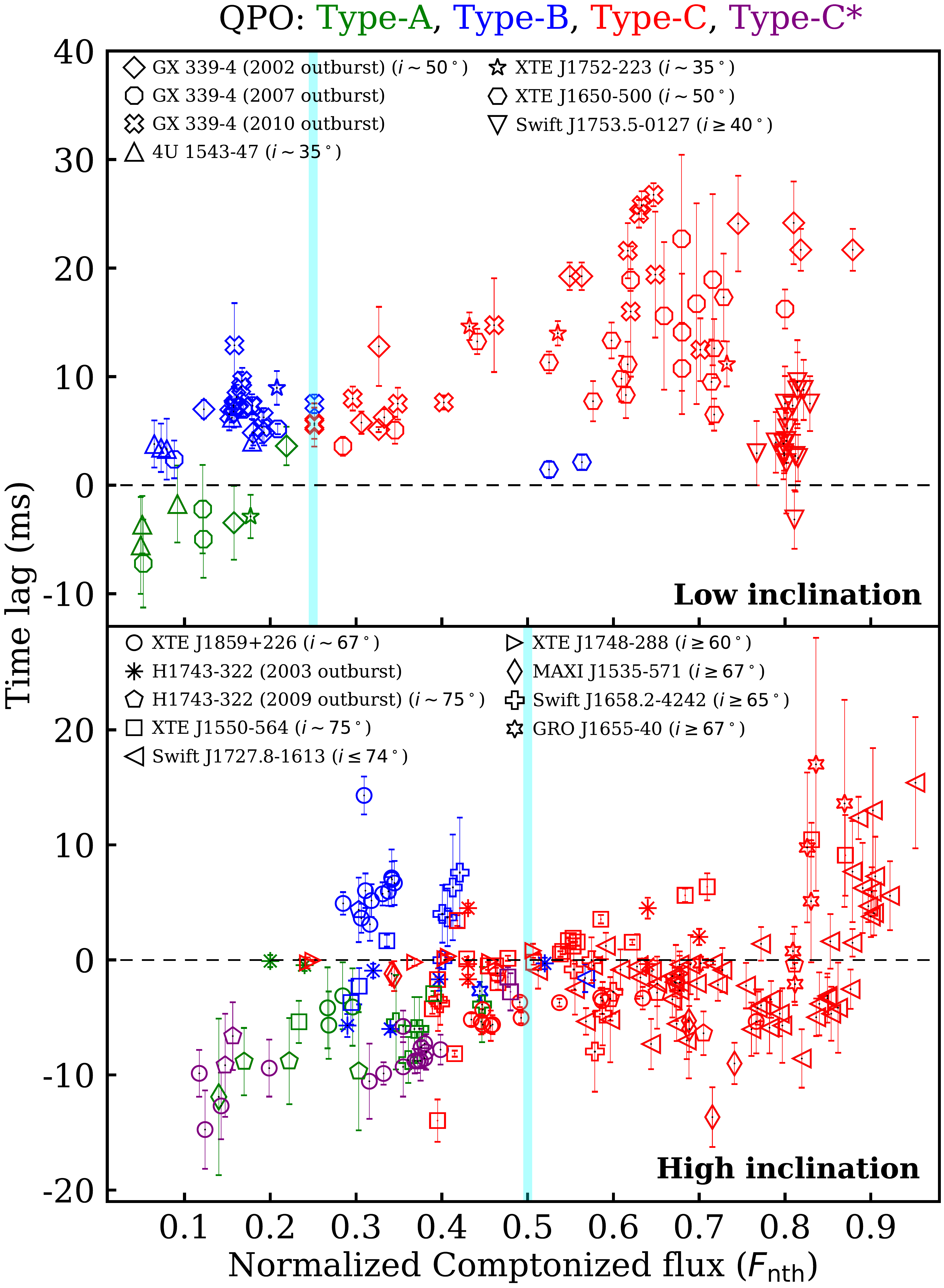}
    \end{center}
    \caption{Variation of time lag with $F_{\rm nth}$ for (\textit{upper panel}) low inclination sources (GX $339-4$, 2002 outburst: Diamond, 2007 outburst: octagon, 2010 outburst: Cross, 4U $1743-47$: Triangle, XTE J$1752-223$: Asterisk, XTE J$1650-500$: Hexagon, Swift J$1753.5-0127$: Inverted Triangle) and (\textit{lower panel}) high inclination sources (XTE J$1859+226$: Circle, H$1743-322$: Eight pointed asterisk (2003 outburst) and Pentagon (2009 outburst), XTE J$1550-564$: Square, Swift J$1727.8-1613$: Left-caret, XTE J$1748-288$: Right-caret, MAXI J$1535-571$: Small Diamond, Swift J$1658.2-4242$: Plus and GRO J$1655-40$: Six pointed star). Vertical line (cyan) separates HIMS and SIMS in both the panels. Green, blue, red and purple colors represent type-A, type-B, type-C and type-C* QPOs. See the text for details.} 
    \label{fig:fig07}
\end{figure}

In Fig. \ref{fig:fig07}, we examine the possible correlation between $F_{\rm nth}$ and the time lag for type-A, type-B, type-C, and type-C* QPOs across all BH-XRBs under consideration. All together, we have identified $26$ type-A, $54$ type-B, $199$ type-C and $21$ type-C* QPOs (see Table \ref{tab:tab01}). The obtained results for low-inclination and high-inclination sources are presented in upper and lower panels, respectively. For low-inclination sources, type-B and type-C QPOs show positive time lags, while type-A QPOs exhibit negative time lags except for one outlier observed during 2002 outburst of GX 339$-$4. In general, a positive correlation seems to exist between $F_{\rm nth}$ and time lag regardless of the type of QPOs, except for Swift J1753.5$-$0127. For high-inclination sources, type-A, type-C and type-C* QPOs generally exhibit negative time lags, except XTE J$1550-564$, Swift J$1727.8-1613$, 2003 outburst of H$1743-322$ and GRO J$1655-40$, where both positive and negative lags are observed in type-C QPOs. Type-B QPOs demonstrate both positive and negative lag in high inclination sources. In certain high inclination sources such as XTE J$1859+226$, XTE J$1550-564$ and H$1743-322$, we notice that some type-C QPOs with higher $F_{\rm nth}$ show relatively smaller negative time lags compared to few type-C* and type-A QPOs. As the sources move toward SIMS from HIMS (demarcated by vertical cyan line), they show a shift to negative time lags, accompanied by a decrease in $F_{\rm nth}$. Note that type-B QPOs generally show positive time lag ($2-12$ ms) with relatively smaller $F_{\rm nth}$ ($\lesssim 0.4$) regardless of the inclination angle of the sources under consideration except H$1743-322$ (2003 outburst), XTE J$1550-564$, GRO J$1655-40$ and Swift J$1727.8-1613$. We discuss the implications of these findings for examining different disc configurations in \S 6. 

\begin{table*}
       \centering
       \caption{Number of type-C, type-B, type-A, and type-C* QPOs identified at the onset of outbursts in the BH-XRBs under consideration.}
       \renewcommand{\arraystretch}{1.0}
       \begin{tabular}{lccccc}
       \hline
        Source & Outburst & \multicolumn{4}{c}{Number of QPOs}  \\ 
        & (Year) & Type-C & Type-B & Type-A & Type-C* \\
        \hline
        GX $339-4$  & 2002 & $10$ & $5$ & $2$ & - \\
        & 2007 & $10$ & $3$ & $3$ & - \\
                  & 2010 & $13$ & $12$ & - & -\\
                  \hline
        4U $1543-47$& 2004 & - & $5$ & $3$ & - \\
         \hline
        XTE J$1752-223$& 2009  & $3$ & $1$ & $1$ & - \\
         \hline
        XTE J$1650-500$& 2001 & $11$ & $2$ & - & -\\
         \hline
        Swift J$1753.5-0127$& 2005 & $20$ & - & - & -\\
         \hline
        XTE J$1859+226$& 1999 & $14$ & $11$ & $4$ & $17$\\
         \hline
        H$1743-322$    & 2003 & $4$ & $5$ & $2$ & -\\
            & 2009 & $6$ & $2$ & $3$ & $2$\\
        \hline
        XTE J$1550-564$& 1998 & $23$  & $3$  & $2$ & $2$  \\
         \hline
        Swift J$1727.8-1613$& 2023 & $56$  & $1$  & - & -  \\
     \hline
    GRO J$1655-40$ & 2005 & $9$  & $1$  & -  & -  \\
     \hline
    XTE J$1748-288$& 1998  & $7$  & -  & -  & -  \\
     \hline
    MAXI J$1535-571$& 2018 & $6$  & -  & $1$  & -  \\
     \hline
    Swift J$1658.2-4242$&2018 & $7$  & $3$  & $5$ &  - \\
    \hline
       \end{tabular}
       \label{tab:tab01}
\end{table*}

\begin{figure}
    \begin{center}
    \includegraphics[width=\columnwidth]{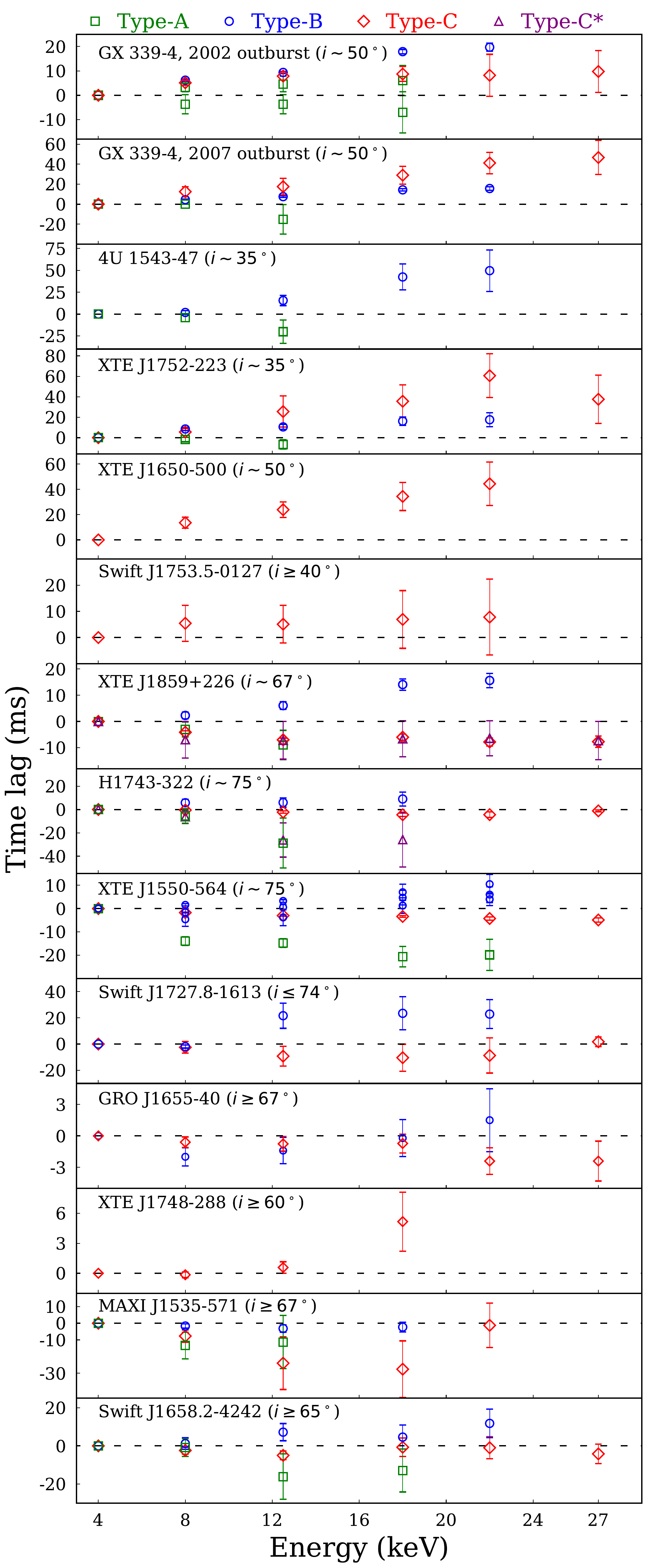}
    \end{center}
    \caption{Evolution of time lag with energy in GX $339-4$ ($i \sim 50^{\circ}$), 2002 and 2007 outburst, 4U $1543-47$ ($i \sim 36^{\circ}$ ), XTE J$1752-223$ ($i \sim 35^{\circ}$ ), XTE J$1650-500$ ($i \ge 47^{\circ}$ ), Swift J$1753.5-0127$ ($i \ge 40^{\circ}$ ), XTE J$1859+226$ ($i \sim 65^{\circ}$), H$1743-322$ ($i \sim 75^{\circ}$ ), XTE J$1550-564$ ($i \sim 74^{\circ}$ ) and Swift J$1727.8-1613$ ($i \le 74^{\circ}$), GRO J$1655-40$ ($i \ge 74^{\circ}$), XTE J$1748-288$ ($i \ge 60^{\circ}$), MAXI J$1535-571$ ($i \ge 67^{\circ}$), Swift J$1658.2-4242$ ($i \ge 65^{\circ}$). Different QPOs are represented by different colored symbols (green square: type-A, blue circle: type-B, red diamond: type-C and purple triangle: type-C*). See the text for details.}
\label{fig:fig08}
\end{figure}

The evolution of time lag as a function of energy for all BH-XRBs is shown in Fig. \ref{fig:fig08}, with a particular focus on QPOs observed during radio flares, as well as additional detection of type-A and type-B QPOs (see Table \ref{tab:tab01}). Square (green), circle (blue), diamond (red) and triangle (purple) denote type-A, type-B, type-C and type-C* QPOs, respectively. We find that type-C QPOs are observed across all energy bands. For low-inclination sources, type-B and type-C QPOs exhibit positive time lags that increase with energy. Type-A QPOs are observed up to $15$ keV in XTE J$1752-223$, 4U $1543-47$ and GX $339-4$ (2007 outburst), and up to $20$ keV in 2002 outburst of GX $339-4$. During the 2002 outburst of GX $339-4$, the positive time lag and the negative lag associated with type-A QPOs exhibit minimal variation across different energy bands. In high-inclination sources, time lags associated with type-C QPOs show minimal variation with energy. Type-A QPOs are detected up to $10-15$ keV and generally exhibit little change in time lag across energies, with the exception of XTE J1550$-$564, which displays a pronounced increase in negative lag with energy.

\begin{figure}
    \centering
    \includegraphics[width=\columnwidth]{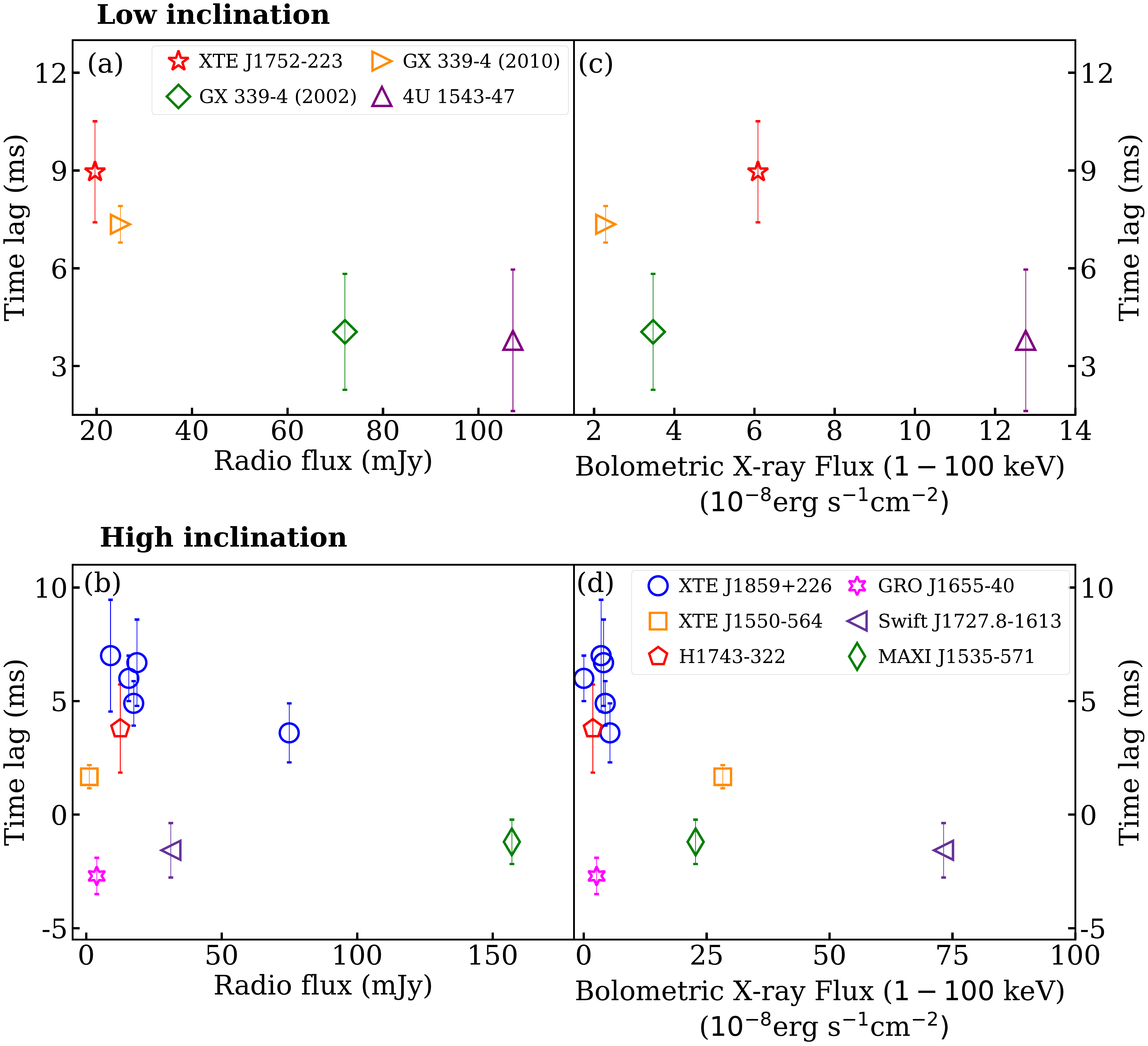}
    \caption{The figure presents the variation of time lag (in milliseconds) with radio flux (in mJy) (in panel (a) and (b)) and bolometric X-ray flux  (in panel (c) and (d)) for different black hole X-ray binaries. Panels (a) and (c) correspond to low-inclination sources (XTE J$1752-223$ (red asterisk), GX $339-4$, $2002$ outburst (green diamond), $2010$ outburst (orange right-caret), and 4U $1543-47$ (purple triangle)), while panels (b) and (d) represent high-inclination sources (XTE J$1859+226$ (blue circle), XTE J$1550-564$ (orange square), H$1743-322$ (red pentagon), GRO J$1655-40$ (magenta six-pointed asterisk), Swift J$1727.8-1613$ (violet left-caret), and MAXI J$1535-571$ (green small-diamond)).}
    \label{fig:fig09}
\end{figure}

Since type-B QPOs often correlate with radio flares, we investigate the relationship between the time lags of type-B QPOs and both radio and X-ray fluxes. Fig. \ref{fig:fig09} presents the variation of time lag with radio flux in panel (a) for low-inclination sources and in panel (b) for high-inclination sources. Panels (c) and (d) display the correlation between time lag and X-ray flux for low- and high-inclination sources, respectively. We normalized the radio flux values to 5 GHz. In low-inclination sources, we find a strong anti-correlation between time lag and radio flux, with a Pearson correlation coefficient $\sim -0.93$. We also identify a weak anti-correlation between time lag and X-ray flux, with a Pearson correlation coefficient $\sim -0.41$. In high-inclination sources, we observe a moderate anti-correlation between radio flux and time lag, yielding a Pearson correlation coefficient $\sim -0.56$ after excluding the source GRO J$1655-40$. Furthermore, we detect a strong anti-correlation between X-ray flux and time lag, with a Pearson correlation coefficient $\sim -0.81$, again excluding GRO J$1655-40$. We also mention that for XTE~J1752$-$223, flare F1 is associated with a type-B QPO, while a type-A QPO is detected prior to flare F2. No QPOs are observed in association with the remaining flares.

\section{Estimation of spin and jet velocity}
\label{sec:section5}

So far, we have explored the evolution and correlations of the spectro-temporal properties during radio flares and spectral state transitions for the BH-XRBs under consideration. We observe an increase in X-ray flux along with the appearance of type-A/type-B QPOs or their absence near the radio flares across all the sources. In this section, we aim to estimate the jet velocity during the radio flares ($i.e.$, transient jets) in SIMS. To do so, we first calculate the mass accretion rate, which is influenced by radiative efficiency --- a factor directly tied to the spin of the sources. In the next sub-section, we present the spin estimates derived using the continuum-fitting method for nine sources, while for the remaining four, we rely on previously reported spin values, as the continuum-fitting method could not be applied (see Section \ref{subsec:subsection3.2}).

\subsection{Estimation of Spin}
\label{sub:subsection5.1}

\begin{table*}
	\centering
	\caption{Fitted parameters using model \texttt{kerrbb2} with $90\%$ confidence in GX $339-4$, 4U $1543-47$,  XTE J$1752-223$, XTE J$1650-500$, XTE J$1859+226$, H$1743-322$,  XTE J$1550-564$ and GRO J$1655-40$, XTE J$1748-288$ and their corresponding radiative efficiency of the accretion disc $\eta_{acc}$. For Swift J$1753.5-0127$, Swift J$1727.8-1613$, Swift J$1658.2-4242$ and MAXI J$1535-571$, the continuum fitting method cannot be applied, and the spin values are taken from the literature.}
	\renewcommand{\arraystretch}{1.5}
	\resizebox{\textwidth}{!}{%
		\begin{tabular}{l c c c c c c c c c c}
			\hline
			\hline
			Source &Date&$D$ &$i$ &$M_{\text{BH}}$ & a$_{\text{k}}$ & $\dot{\text{M}}_{acc}^{\boxplus}$& $\eta_{acc}$&References\\
			&(MJD)& (kpc) & ($^\circ$) &(M$_{\odot}$) &  & ($\dot{\text{M}}_{Edd}$)& &\\\hline
			GX $339-4$& $52660.64$& $8.4$& $50$& $9.4$& $0.53\pm0.02$& $0.12^{\dagger}$&$0.08-0.09$ &[1], [2], [3], [4], [5] \\
			& $52667.02$& & & & $0.60_{-0.05}^{+0.03}$&$0.11_{-0.01}^{+0.01}$& & \\
			& $52671.25$& & & & $0.58\pm0.03$& $0.10\pm0.01$  & &\\
			4U $1543-47$& $52468.72$&$7.5$ &$35$ &$9.4$ & $0.44\pm0.05$& $0.16\pm0.01$  &$0.07-0.09$ &[6], [7]\\
			& $52469.23$& & & & $0.43_{-0.05}^{+0.06}$& $0.16\pm0.01$& & \\
			& $52473.17$& & & & $0.52_{-0.08}^{+0.06}$&$0.10\pm0.01$& & \\
			XTE J$1752-223$& $55252.59$&$7.1$ &$35$ &$12$ & $0.52_{-0.09}^{+0.07}$& $0.08^\dagger$  &$0.07-0.10$ &[8], [9], [10], [11], [12]\\
			& $55256.74$& & & & $0.62_{-0.06}^{+0.02}$& $0.06\pm0.01$& & \\
			& $55260.80$& & & & $0.46_{-0.09}^{+0.11}$&$0.08\pm0.01$& & \\
			XTE J$1650-500$& $52174.29$&$2.6$ &$50$ &$5.1$ & $0.63_{-0.07}^{+0.06}$& $0.10^\dagger$&$0.09-0.11$ &[13], [14], [15], [16]\\
			& $52217.47$&  &  &  & $0.65\pm0.05$& $0.03^\dagger$  & & \\
			& $52352.65$& & & & $0.62_{-0.07}^{+0.10}$& $0.04\pm0.01$& & \\
			XTE J$1859+226$& $51502.25$&$6$ &$67$ &$7.8$ & $0.34_{-0.04}^{+0.03}$& $0.13\pm0.01$&$0.07-0.08$ &[12], [17], [18], [19], [20]\\
			& $51504.32$&  &  &  & $0.32_{-0.04}^{+0.05}$& $0.13\pm0.01$  & & \\
			& $51520.08$& & & & $0.34_{-0.06}^{+0.07}$& $0.10\pm0.01$& & \\
			H$1743-322$& $54997.26$&$8.5$ &$75$ &$11.2$ & $0.36_{-0.06}^{+0.05}$& $0.17\pm0.01$&$0.07-0.08$ &[21], [22]\\
			& $54999.74$&  &  &  & $0.43_{-0.09}^{+0.06}$& $0.14\pm0.01$  & & \\
			& $55220.36$& & & & $0.39_{-0.15}^{+0.14}$& $0.13\pm0.03$& & \\
			XTE J$1550-564$& $51266.87$&$4.4$ &$75$ &$9.1$ & $0.35_{0.03}^{0.05}$& $0.16\pm0.01$&$0.07-0.08$ &[23], [24], [25]\\
			& $51270.74$& & & & $0.38_{-0.05}^{+0.03}$& $0.10\pm0.01$& & \\
			& $51274.47$& & & & $0.42_{-0.05}^{+0.03}$& $0.10\pm0.01$& & \\
            GRO J$1655-40$& $53449.84$&$3.2$ &$65$ &$6.0$ & $0.82_{-0.003}^{+0.010}$& $0.12\pm0.01$&$0.11-0.13$ &[26], [27], [28], [29], [30]\\
			& $53616.79$& & & & $0.78\pm0.01$& $0.04\pm0.01$& & \\
			& $53619.87$& & & & $0.76_{-0.04}^{+0.01}$& $0.04\pm0.01$& & \\
            XTE J$1748-288$& $50986.93$&$8.7$ &$65$ &$7.6$ & $0.80\pm0.01$& $0.23\pm0.01$&$0.12-0.13$ &[12], [31]\\
			& $50991.48$& & & & $0.81\pm0.01$& $0.09\pm0.01$& & \\
			& $50996.75$& & & & $0.83\pm0.02$& $0.07\pm0.01$& & \\
            
			\hline  
			Swift J$1753.5-0127$ &- &$7.1$ &$\ge 40$ &$7.4$&$0.989_{-0.035}^{+0.007}$& - &$0.19-0.29$ & [32], [33], [34], [35] \\
			Swift J$1727.8-1613$ &- &$3.7$ &$\le74$ &$3.1$&$0.98_{-0.07}^{+0.02}$&-& $0.16-0.30$ & [36], [37], [38], [39] \\
            MAXI J$1535-571$ &- &$3.7$ &$\ge67$ &$3.1$&$ 0.992\pm0.001$&-& $0.26-0.28$ & [40], [41], [42], [43], [44] \\
            Swift J$1658.2-4242$ &- &$6.3$ &$\ge65$ &$14$&$\ge0.96$&-& $\ge0.20$ &[45], [46]  \\
            
			\hline

		\end{tabular}
	}
	\label{tab:tab02}
	$^{\boxplus}\dot{M}_{Edd}=1.47\times10^{18}$ $M_{\rm BH}/M_\odot$ erg s$^{-1}$.$^\dagger$Error is insignificant upto two decimal places.\\
	Reference: [1] \cite{Heida-etal-2017}, [2] \cite{Furst-etal-2015}, [3] \cite{Parker-etal-2016}, [4] \cite{Zdiarsky-etal-2019}, [5] \cite{Shidatsu-etal-2011}, [6] \cite{Dong-etal-2020}, [7] \cite{Chen-Wang-2024},  [8] \cite{Miller-etal-2011}, [9] \cite{Shaposhnikov-etal-2010}, [10] \cite{Garcia-etal-2018}, [11] \cite{Debnath-etal-2021}, [12] \cite{Abdulghani-etal-2024}, [13] \cite{Slany-etal-2008}, [14] \cite{Orosz-etal-2004}, [15] \cite{Homan-etal-2006}, [16] \cite{Mondal-etal-2010}, [17] \cite{Hynes-etal-2002}, [18] \cite{Motta-etal-2022}, [19] \cite{Corral-Santana-etal-2013}, [20] \cite{Rizo-etal-2022}, [21] \cite{Steiner-etal-2012}, [22] \cite{Molla-etal-2017}, [23] \cite{Jonker-etal-2010}, [24] \cite{Jonker-etal-2004}, [25] \cite{Orosz-etal-2011}, [26] \cite{Ponti-etal-2012}, [27] \cite{Shafee-etal-2006}, [28] \cite{Greene-etal-2001}, [29] \cite{Kuulkers-etal-2000}, [30] \cite{Hjellming-etal-1995}, [31] \cite{Eijnden-etal-2017}, [32] \cite{Shaw-etal-2016}, [33] \cite{Gandhi-etal-2019}, [34] \cite{Neustroev-etal-2014}, [35] \cite{Draghis-etal-2024}, [36] \cite{MataSanchez-etal-2024}, [37] \cite{Peng-etal-2024}, [38] \cite{Wood-etal-2024}, [39]  \cite{Yu-etal-2024}, [40] \cite{Sreehari-etal-2019}, [41] \cite{Sridhar-etal-2019}, [42] \cite{Liu-etal-2022}, [43] \cite{Chauhan-etal-2019}, [44] \cite{Miller-etal-2018}, [45] \cite{Xu-etal-2018}, [46] \cite{Mondal-etal-2023}.
\end{table*}

We constrain the spin ($a_{\rm k}$) of nine BH-XRBs by fixing their mass ($M_{\rm BH}$), inclination ($i$), and distance ($D$) as described in Section \ref{subsec:subsection3.2}. For each source, we select three observations that meet the continuum fitting conditions, namely, $\text{f\_scat} < 25\%$ and $L_{\rm bol}/L_{\rm Edd} < 0.3$ \cite[]{Steiner-etal-2011}, where $L_{\rm Edd}$ represents the Eddington luminosity and $L_{\rm bol}$ refers to the bolometric luminosity. The estimated spin ($a_{\rm k}$) and accretion rate ($\dot M$) of each source is presented in Table \ref{tab:tab02}. It is worth mentioning that the estimated $a_{\rm k}$ values for all sources are consistent with their previous estimates (see references in Table \ref{tab:tab02}). Moreover, a good agreement is observed for the mass accretion rate $\dot M$ as obtained using model combination \texttt{tbabs}$\times$(\texttt{thcomp}$\otimes$\texttt{diskbb}). For Swift J1753.5$-$0127 and Swift J1727.8$-$1613, the spin values ($a_{\rm k}$) are constrained using reflection modelling as $0.989^{+0.007}_{-0.035}$ \cite[]{Draghis-etal-2024} and $0.98^{+0.02}_{-0.07}$ \cite[]{Peng-etal-2024}, respectively. For MAXI J1535$-$571 and Swift J1658.2$-$4242, spin values were adopted from the literature. The continuum-fitting method could not be applied to Swift J1658.2$-$4242, as it was not observed in the HSS. In the case of MAXI J1535$-$571, reliable spin estimation was not possible due to the poor quality of low-energy data \cite[]{Chen-etal-2020}.
Subsequently, we calculate the radiative efficiency of accretion disc ($\eta_{acc}$) using source spin values $a_{\rm k}$ following \cite{Thorne-1974,Hobson-etal-2006} and present in Table \ref{tab:tab02}.

\subsection{Estimation of Jet Velocity}
\label{sub:subsection5.2}

Following \cite{Longair-2011}, we estimate the jet velocity by applying minimum energy condition which implies that the minimum energy contained in the blob of ejected plasma is given by,
\begin{equation}
W_{\rm min}=3\times10^6 ~ \eta^{4/7}V^{3/7}\nu^{2/7}S_{\nu}^{4/7}D^{8/7}~ \text{Joule},
\end{equation}
where $S_\nu$ is the observed radio flux at radio frequency $\nu$, $D$ is the distance of the source and $V$ is the volume of the spherical plasma blob. Here, $\eta$ denotes the relativistic correction factor assumed to be unity \cite[]{Brocksopp-etal-2002, Nandi-etal-2018}. Considering $V=(\beta ct)^3$, where $t$ is the rise-time of the radio event and $\beta$ is the jet velocity in the unit of $c$, the observed jet power is estimated as, 
\begin{equation}
    \label{eq:eqn7}
    \begin{aligned}
	L_{\rm jet}= \frac{W_{\rm min}}{t}= & ~3\times10^{33}\left(\frac{S_\nu}{\rm mJy} \right)^{4/7}\left(\frac{D}{\rm kpc}\right)^{8/7}\\
	&\left(\frac{\nu}{\rm GHz}\right)^{2/7}\left(\frac{t}{\rm s}\right)^{2/7}\beta^{9/7}~\text{erg s$^{-1}$}.
    \end{aligned}
\end{equation}
We use equation (\ref{eq:eqn1}) to normalize the radio flux at frequency $\nu=5$ GHz. Following \cite{Fender-2001, Miller-etal-2006}, we apply Doppler correction and obtain the intrinsic radio flux ($S_{\nu}^{\rm int}$) using observed radio flux ($S_{\nu}$) as $S_{\nu}^{\rm int}=\delta^{\alpha-k}S_{\nu}$, where $\nu=\delta \times \nu^{\rm int}$, $\delta^{-1}=\gamma(1-\beta \cos i)$, $\delta$ being the Doppler factor and $\gamma~(=1/\sqrt{1-\beta^2})$ is the Lorentz factor. We assume the radio flux to follow a power law $S_{\nu}=\nu^{\alpha}$, where $t$ transforms as $t=t^{\rm int}/\delta$ for simplicity. The parameter $k$ accounts for the characteristics of the ejecta, where 
$k=2$ and $k=3$ refer steady continuous jet and discrete jet, respectively \cite[]{Rybicki-Lightman-1979}. With this, the intrinsic jet power is calculated as,
\begin{equation}
\label{eq:eqn8}
L^{\rm int}_{\rm jet} = \delta^{4(\alpha-k)/7} \times L_{\rm jet} = \left[ \gamma(1-\beta \cos i) \right]^{4(k-\alpha)/7} \times L_{\rm jet}.
\end{equation}

During accretion, a part of the inflowing matter may deflect along the rotation axis of the black hole producing bipolar jets. For a given mass accretion rate ($\dot M$), we compute the jet kinetic power as,
\begin{equation}
\label{eq:eqn9}
L^{\rm est}_{\rm jet}=\frac{1}{2}~\eta_{\rm jet} \times \varepsilon \times \dot{M} \times c^2,
\end{equation}
where $\eta_{\rm jet}$ is the jet radiative efficiency factor assumed as $\sim 0.1$ \cite[]{Fender-2001b} and $\varepsilon$ accounts the fraction of energy transferred to jets from the disc \cite[]{Chakraborti-1999,Das-etal-2001,Aktar-etal-2019}.

Given the unabsorbed X-ray flux ($F_{\rm x}$) of a source with distance $D$, the accretion rate can be calculated using the X-ray luminosity $L_{\rm x} ~(=4 \pi D^2 F_{\rm x})$ as,
\begin{equation}
    {\dot M} = 8.73 \times 10^{-17} \times \eta_{\rm acc} \left( \frac{F_{\rm x} D^2}{c^2} \right) \left( \frac{M_{\rm BH}}{M_\odot}\right)^{-1} ~ {\dot M}_{\rm Edd},
\end{equation}
where ${\dot M}_{\rm Edd}=1.47 \times 10^{18} \left( M_{\rm BH}/M_\odot\right)$ g s$^{-1}$. In this study, we focus on observations taken at the peak bolometric X-ray flux or during type-A/type-B QPOs in SIMS, particularly around the radio flares, while for HIMS, the X-ray flux is determined using the closest available X-ray observation relative to the radio flare. The rise time $t$ is estimated as the interval between the X-ray observation used to calculate $\dot M$ and the observation with radio flare. Using equation (\ref{eq:eqn7}) in equation (\ref{eq:eqn8}) and then equating with equation (\ref{eq:eqn9}), we estimate the jet velocity $\beta$ in unit of $c$. Following \cite{Fender-etal-2004, Fender-etal-2009}, we adopt $k=2$ during the HIMS for continuous jets, and $k=3$ during the SIMS, when jets are discrete.

\begin{figure}
    \begin{center}
    \includegraphics[width=\columnwidth]{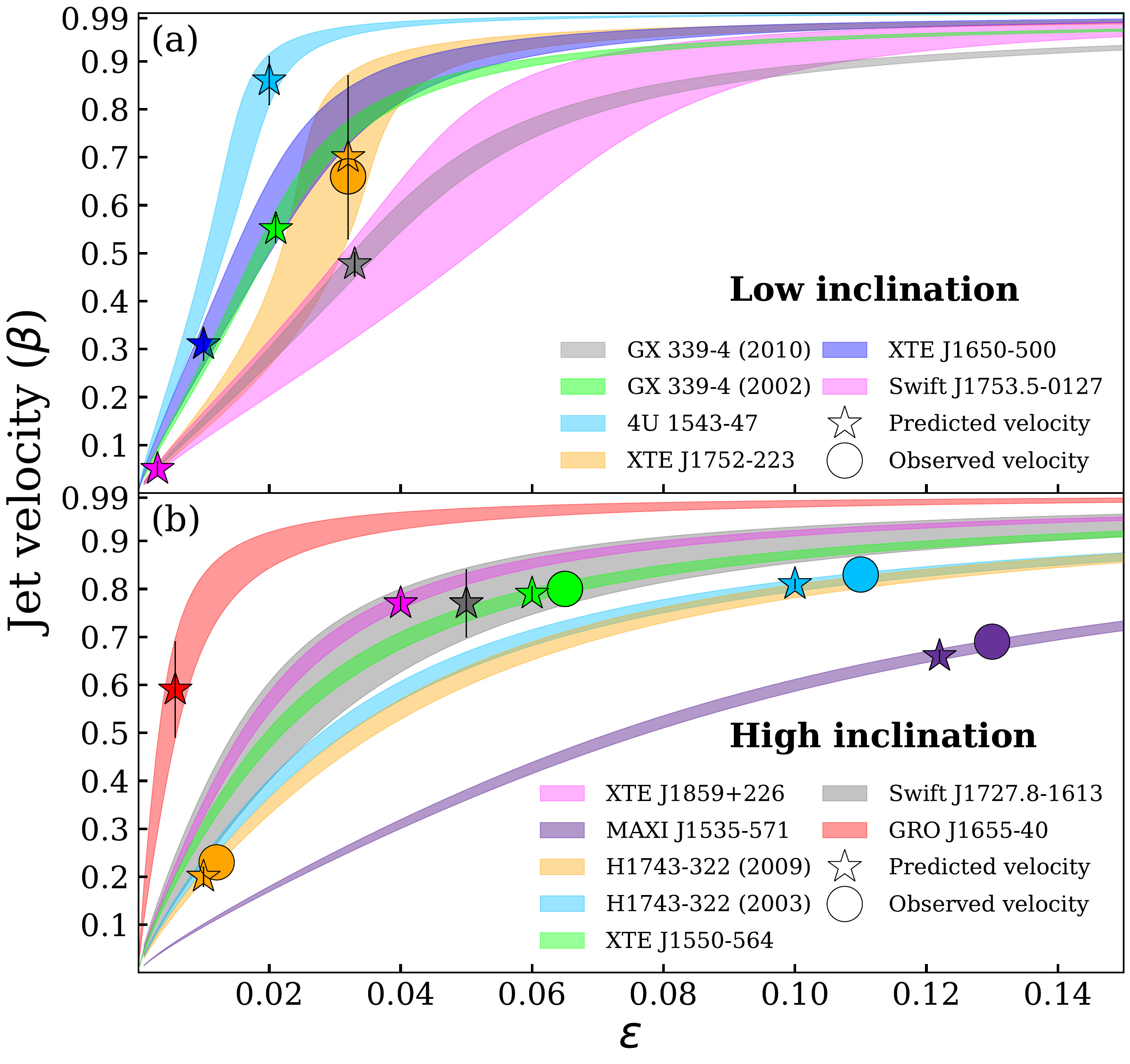}        
    \end{center}
    \caption{Variation of jet velocity (shaded region) with $\varepsilon$ (accounts the fraction of energy transferred to jets from disc) for (a) low-inclination sources such as, GX $339-4$ (green for 2002 outburst and grey for 2010 outburst), 4U $1543-47$ (skyblue), XTE J$1752-223$ (orange), XTE J$1650-500$ (blue), Swift J$1753.5-0127$ (pink) and (b) high-inclination sources such as, XTE J$1859+226$ (pink), MAXI J$1535-571$ (violet), H$1743-322$, $2009$ outburst (orange), H$1743-322$, $2003$ outburst (skyblue), XTE J$1550-564$ (green), Swift J$1727.8-0127$ (grey), GRO J$1655-40$ (red). The values of radiative efficiency of disc $\eta_{\rm acc}$ are taken from Table 1. Asterisks represent the jet velocity $\beta$ predicted from our calculations and circles represent the velocity reported earlier.} 
    \label{fig:fig10}
\end{figure}

In Fig. \ref{fig:fig10}, we depict the variation of jet velocity ($\beta$) with $\varepsilon$ for BH-XRBs under consideration. Panel (a) shows the results for sources with low-inclination, including GX 339$-$4 (2002 and 2010 outbursts), 4U 1543$-$47, XTE J1752$-$223, XTE J1650$-$500, and Swift J1753.5$-$0127. Panel (b) displays the results for sources with high-inclination, such as  XTE J$1859+226$, MAXI J$1535-571$, H$1743-322$, (2003 and 2009 outbursts), XTE J$1550-564$, Swift J$1727.8-0127$ and GRO J$1655-40$. The shaded regions denote the range of velocities resulting due to the uncertainty in estimating $\eta_{\rm acc}$ that largely depends on the black hole spin ($a_{\rm k}$) estimate (see Table \ref{tab:tab02}). Note that jet velocity couldn't be reliably estimated for Swift J1658.2$-$4242 and XTE J1748$-$288 due to large uncertainties arising from the observational gap between X-ray and radio data. The radiative efficiency of the disc is expressed $\eta_{\rm acc} = 1 - \mathcal{E}_{\rm ISCO}$ \cite[]{Thorne-1974, Hobson-etal-2006, Shapiro-Teukolsky-1983}, where 
the specific energy ($\mathcal{E}_{\rm ISCO}$) at the innermost stable circular orbit ($R_{\rm ISCO}$) is given by, $\mathcal{E}_{\rm ISCO}  = \left( 1 - \frac{2M}{R_{\rm ISCO}} \pm a_k \sqrt{\frac{M}{R_{\rm ISCO}^3}}\right) \left( 1 - \frac{3M}{R_{\rm ISCO}} \pm 2a_{\rm k} \sqrt{\frac{M}{R_{\rm ISCO}^2}} \right)^{-1/2}.$ For a Kerr black hole, $R_{\rm ISCO}$ is expressed as \cite[]{Bardeen-etal-1972,Shakura-Sunyeav-1973}, $R_{\rm ISCO} = M \left( 3 + Z_2  \mp \sqrt{(3 - Z_1)(3 + Z_1 + 2Z_2)} \right)$, where\\ $Z_1 = 1 + \left(1 - (a_{\rm k}/M)^2\right)^{1/3} \left[ \left(1 + a_{\rm k}/M\right)^{1/3} + \left(1 - a_{\rm k}/M\right)^{1/3} \right]$, $Z_2 = \sqrt{3(a_{\rm k}/M)^2 + Z_1^2}$ and $M$ is the mass of the black hole. Here, the negative sign corresponds to prograde orbits, while the positive sign applies to retrograde orbits. We observe that $\beta$ increases with $\varepsilon$ for all sources. For a given $\varepsilon$, $\beta$ has both upper and lower limits: the upper limit corresponds to the minimum estimate of $a_{\rm k}$, while the lower limit corresponds to the maximum estimate of $a_{\rm k}$. These findings suggest that the impact of black hole spin $a_{\rm k}$ on determining the jet velocity $\beta$ appears to be minimal.

Meanwhile, the jet velocity during the radio flare has been reported for few sources. For H$1743-322$, the jet velocities during the $2003$ and $2009$ outbursts were measured as $\beta \sim 0.8$ and $\beta \sim 0.2$, respectively, based on the proper motion of the jet ejecta \cite[]{Corbel-etal-2005,Miller-etal-2012}. Using these, we estimate $\varepsilon \sim 0.10$ for $2003$ outburst and $0.01$ for $2009$ outburst. For the $1998$ outburst of XTE J1550$-$564, \cite{Hannikainen-etal-2009} reported the jet velocity $\beta \ge 0.8$ that holds for $\varepsilon \ge 0.06$. For MAXI J$1535-571$, $\beta$ is reported as $\sim 0.69$  \cite[]{Russell-etal-2019}, that corresponds to $\varepsilon\ge0.12$. In the case of F4 in XTE J$1752-223$, the jet velocity was estimated as $\beta \ge 0.66$ \cite[]{Miller-etal-2011} yielding $\varepsilon \sim 0.03$. The open circles in Fig. \ref{fig:fig10} represent the observed values of $\beta$. These findings clearly indicate that the observed $\beta$ is resulted from approximately $1-12$\% of the accreting matter being ejected from the disc as jets in BH-XRBs. For these sources, the predicted values of $\varepsilon$ are in good agreement with the difference between the normalized Comptonized fluxes ($F_{\rm nth}$ ) of successive observations during radio flares. Considering this, we estimate the efficiency $\varepsilon$ during the radio flares and predict the jet velocity $\beta$ for the remaining sources, which are marked using asterisk in Fig. \ref{fig:fig10} for each sources, and the results are summarized in Table \ref{tab:tab03}.

\begin{table}
    \caption{Estimated jet velocities ($\beta$) during radio flares in the BH-XRBs under consideration. For sources exhibiting multiple radio flares, the resulting $\beta$ values vary accordingly.}
	\renewcommand{\arraystretch}{1.5}
	\resizebox{\columnwidth}{!}{%
		\begin{tabular}{l c c c c}	
		\cline{1-5}
		Source & \multicolumn{2}{|c}{Date$^{\dagger}$}& $\varepsilon$ & $\beta$\\
				& Before flare & After flare &    & \\
				\cline{1-5}
		GX $339-4$&$52406.0$ &$52408.0$ &$0.021 $ &$0.52-0.58 $\\
            GX $339-4$&$55310.0$ &$55314.1$ &$0.033 $ &$0.45-0.50 $\\
				\hline
		4U $1743-47$&$52443.2$ &$52445.5$ &$0.02 $ &$0.81-0.91 $\\
				\hline
		XTE J$1752-223$ (F1) &$55218.1$ & $55218.8$ &$0.061 $ &$0.96-0.98$\\
		XTE J$1752-223$ (F2)  &$55220.37$ &$55222.92$ &$0.061 $ &$0.37-0.75$\\
		XTE J$1752-223$ (F3)  &$55221.08$ &$55223.86$ &$0.007 $ &$0.33-0.61$\\
		XTE J$1752-223$ (F4)  &$55231.61$ &$55238.93$ &$0.032 $ &$0.53-0.87$\\
		XTE J$1752-223$ (F5)  &$55249.65$ & $55250.00$ &$0.012 $ &$0.26-0.42$\\
		XTE J$1752-223$ (F6)  &$55258.71$ &$55260.83$ &$0.005 $ &$0.13-0.17$\\
		XTE J$1752-223$ (F7)  &$55278.58$&$55278.96$&$0.005 $ &$0.14-0.18$\\
				\hline
            XTE J$1650-500$   &$52160.4$ & $52161.1$&$ 0.012$ &$0.23-0.34$\\
				\hline
		Swift J$1753.5-0127$ &$53560.5$ &$53567.0$ &$0.003 $ &$0.03-0.06$\\
				\hline
		XTE J$1859+226$ (HIMS)&$51464.1$ &$51464.6$ &$0.002 $ &$0.08-0.09$\\
		XTE J$1859+226$ (F1) &$51467.9$ &$51468.5$ &$0.040 $ &$0.75-0.78$\\
		XTE J$1859+226$ (F2) &$51474.3$&$51474.8$ &$0.030 $ &$0.72-0.76$\\
		XTE J$1859+226$ (F3)  &$51477.1$ &$51478.8$ &$0.030 $ &$0.65-0.69$\\
		XTE J$1859+226$ (F4) &$51483.9$ &$51484.07$ & $0.020 $ &$0.65-0.69$\\
				\hline
		H$1743-322$ (2009) &$54987.2$ &$54989.1$  &$0.010 $ &$0.18-0.22$\\
		H$1743-322$ (2003) &$52735.7$ &$52737.5 $ &$0.100 $ &$0.80-0.82$\\
				\hline
		XTE J$1550-564$   &$51075.9$ &$51078.1$ &$0.061 $ &$0.78-0.81$\\
				\hline
		Swift J$1727.8-1613$ &$60206.1$ &$60211.1$ &$0.050 $ &$0.70-0.84$\\
		Swift J$1727.8-1613$ &$60221.8$ &$60222.1$ &$0.020 $ &$0.55-0.73$\\
				\hline
            GRO J$1655-40$ &$53446.89$ &$53448.01$ &$0.006 $ &$0.49-0.69$\\
                \hline
            MAXI J$1535-571$ &$58019.1$ &$58019.9$ &$0.122 $ &$0.65-0.67$\\
                \hline
			\end{tabular}
			\label{tab:tab03}
		}
    $^{\dagger}$Successive X-ray observations during radio flare are used to estimate $\varepsilon$.
\end{table}

\begin{figure}
    \begin{center}
    \includegraphics[width=\columnwidth]{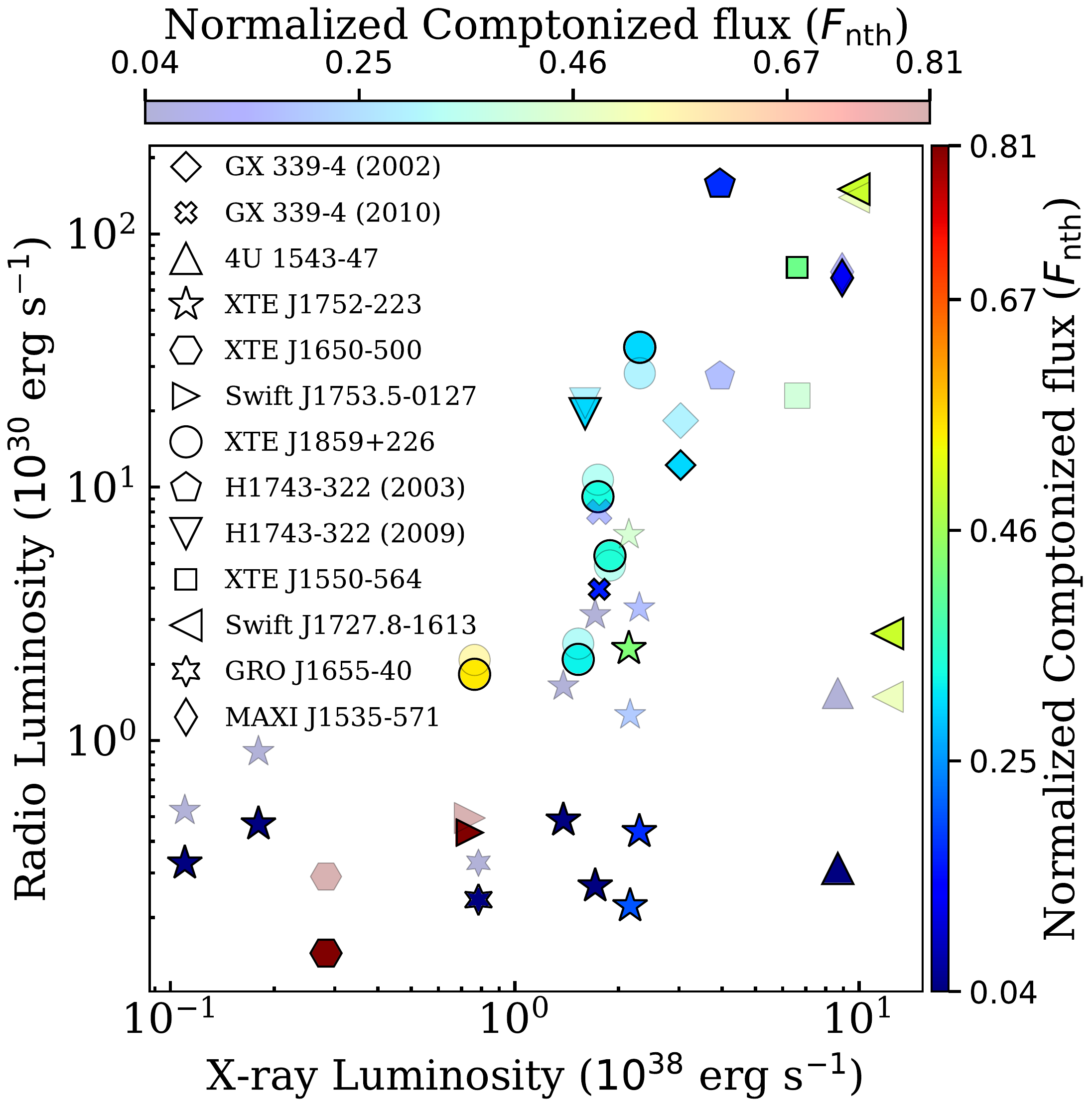}
    \end{center}
    \caption{Variation of intrinsic radio luminosity $L^{\rm int}_{\rm R}$ (opaque) and $L_{\rm R}$ (semi-transparent) observed radio luminosity with X-ray luminosity for GX 339$-$4 (2002 and 2010 outbursts denoted with diamond and cross), 4U 1543$-$47 (triangle), XTE J1752$-$223 (asterisk), XTE J1650$-$500 (hexagon), Swift J1753.5$-$0127 (right-caret), XTE J1859$+$226 (circle), H1743$-$322 (2003 and 2009 outbursts are shown with pentagon and inverted triangle), XTE J1550$-$564 (square), Swift J1727.8$-$1613 (left-caret), GRO J$1655-40$ (six-pointed stars) and MAXI J$1535-571$ (small diamond). Colorbars at the top and right of the figure denote the range of $F_{\rm nth}$ obtained from spectral modelling. See the text for details.} 
    \label{fig:fig11}
\end{figure}

Since the observed radio luminosity $L_{\rm R}~(=\nu S_\nu D^2$) is likely to be influenced by Doppler boosting, we investigate the relationship between $L_{\rm R}$ and the intrinsic radio luminosity $L^{\rm int}_{\rm R}~ (=\delta^{\alpha -k -1} L_{\rm R}$) to assess the impact of Doppler correction on the measured values. In Fig. \ref{fig:fig11}, we present the variations of $L_{\rm R}$ and $L^{\rm int}_{\rm R}$ with X-ray luminosity ($L_{\rm bol}$). Results obtained for different sources are represented by distinct symbols: GX 339$-$4 (2002 and 2010 outbursts are denoted with diamond and cross), 4U 1543$-$47 (triangle), XTE J1752$-$223 (asterisk), XTE J1650$-$500 (hexagon), Swift J1753.5$-$0127 (right-caret), XTE J1859$+$226 (circle), H1743$-$322 (2003 and 2009 outbursts are shown with pentagon and inverted triangle), XTE J1550$-$564 (square), Swift J1727.8$-$1613 (left-caret), GRO J$1655-40$ (six-pointed stars) and MAXI J$1535-571$ (small diamond). Opaque and semi-transparent symbols represent $L^{\rm int}_{\rm R}$ and $L_{\rm R}$, respectively, with colors indicating the range of normalized Comptonized flux ($F_{\rm nth}$), as shown by the colorbar at the top and right of the figure. All sources exhibit a normalized Comptonized flux value of $F_{\rm nth}>0.5$ in HIMS and $F_{\rm nth}\leq 0.5$ in SIMS. Sources in HIMS show lower radio and X-ray luminosities compared to those in SIMS. The difference between $L_{\rm R}$ and $L^{\rm int}_{\rm R}$ possibly suggests the influence of relativistic beaming on jet emissions. Furthermore, we observe that both $L_{\rm R}$ and $L^{\rm int}_{\rm R}$ increase with the increase of X-ray luminosity. However, in XTE J1752$-$223, corresponding to F6 and F7, the source is in the soft state, where both X-ray and radio luminosities as well as the normalized Comptonized flux are lower. It is important to note that for low-inclination sources, the observed luminosity ($L_{\rm R}$) tends to be higher than the intrinsic luminosity ($L^{\rm int}_{\rm R}$). This pattern also holds for high-inclination sources when the jet velocity remains in the range $\beta \sim 0.13-0.65$. However, for highly inclined sources ($i \gtrsim 60^\circ$) with higher jet velocities, the intrinsic luminosity ($L^{\rm int}_{\rm R}$) exceeds the observed luminosity ($L_{\rm R}$).

\section{Discussion and Conclusion}
\label{sec:section6}

In this work, we perform a comprehensive spectro-temporal analyses of thirteen BH-XRBs during their entire outbursts using \textit{RXTE}, \textit{Insight/HXMT} and \textit{AstroSat} observations. We examine the spectro-temporal properties during the radio flares to investigate the disc-jet connection. Further, we focus on estimating the accretion rate ($\dot{m}$) and black hole spin (a$_k$), while computing the velocity ($\beta$) of the plasma ejected in the form of jets.

\subsection{Spectro-temporal Properties to Probe Disc-Jet Connection}
\label{sec:6.1}

The transition from HIMS to SIMS is characterized by an increase in X-ray flux, which coincides with the onset of radio flares, as shown in Fig. \ref{fig:fig01}. This transition is also linked to the appearance of type-A and type-B QPOs near radio flares in both high- and low-inclination sources (see Fig. \ref{fig:fig02} and Fig. \ref{fig:fig04}). Similar coincidence of X-ray peak and radio flare is also observed in SIMS, both in presence and absence of QPOs (see Fig. \ref{fig:fig06}). In high-inclination sources, such as XTE J1859$+$226 and H1743$-$322, type-B QPOs emerge near radio flares, displaying a positive time lag, in contrast to the negative time lag observed with type-C QPOs (see Fig. \ref{fig:fig07}). Similarly, in XTE J$1550-564$, H$1743-322$ (2003 outburst), GRO J$1655-40$ and Swift J$1727.8-1613$, type-A and type-B QPOs are detected near radio flares with negative time lag, while type-C QPOs show positive time lag with relatively higher $F_{\rm nth}$. For low-inclination sources, such as GX 339$-$4, 4U 1543$-$47, and XTE J1752$-$223, type-B QPOs are observed near radio flares with positive time lags, while XTE J1650$-$500 lacks concurrent radio observations during the occurrence of type-A or type-B QPOs (see Fig. \ref{fig:fig06}). During the failed outburst of Swift J1753.5$-$0127, only type-C QPOs are detected with positive time lag.

We observe that higher $\rm rms_{\rm QPO}\%$ in type-C QPOs is correlated with normalized Comptonized flux ($F_{\rm nth}$) in all sources (see Fig. \ref{fig:fig06}). These findings indicate that type-C QPOs are possibly originated from the disc-corona \cite[see also][]{Titarchuk-Fiorito-2004, Ingram-etal-2009, Nandi-etal-2012, Iyer-etal-2015, Motta-etal-2015, Nandi-etal-2024}. Meanwhile, the origin of type-B QPOs has also been attributed to the corona \cite[]{Gao-etal-2014, Motta-etal-2015, Garcia-etal-2021, Belloni-etal-2020, Zhang-etal-2023}, although \cite{Fender-etal-2009, Radhika-Nandi-2014, Radhika-etal-2016, Harikrishna-Sriram-2022, Zhang-etal-2023} suggest a potential link between type-B QPOs and the radio jets. This emphasizes the crucial role of the corona in driving both QPO variability and jet ejections, and hence, we investigate how the time-lag varies with $F_{\rm nth}$ (see Fig. \ref{fig:fig07}). In low-inclination sources, type-C and type-B QPOs generally show positive lags, whereas type-A QPOs exhibit both positive and negative lags. In high-inclination sources, type-C QPOs display both positive and negative lags depending on the Comptonized emission, while Type-C* and type-A QPOs predominantly exhibit negative lags. Type-B QPOs having similar $F_{\rm nth}$ as that of type-A QPO shows both positive and negative lag. Our results align with the finding of \cite{Eijnden-etal-2017} for type-C QPOs and type-A QPOs, which suggests that the time lag of type-C QPOs is influenced by the source inclination, except for Swift J$1727.8-1613$, GRO J$1655-40$, XTE J$1550-564$ and 2003 outburst of H$1743-322$. In contrast, the time lag for type-A QPO appears to be independent of sources inclination. For type-B QPOs, the time lag does show an inclination dependence, particularly low-inclination sources exhibit positive lags, while high-inclination sources display both positive and negative lags.

It is worth mentioning that positive time lags are generally attributed due to the inverse Comptonization of soft photons in the corona \cite[]{Reig-etal-2000}, while negative time lags are associated with multiple physical processes, including Compton down-scattering in the corona \cite[]{Reig-etal-2000}, reprocessing of hard photons in the accretion disc \cite[]{Uttley-etal-2014, Karpouzas-etal-2020}, and gravitational bending \cite[]{Dutta-etal-2016, Chatterjee-etal-2017}. Meanwhile, \cite{Dutta-etal-2016} argued that a larger Comptonizing region reduces QPO frequency, thereby increasing the time lag. The observed rise in positive time lags with increasing Comptonized emissions strongly suggests that the size of the corona significantly influences the time lag characteristics. We find that the positive time lags of type-C QPOs generally increase with energy, while negative time lags show marginal variation (see Fig. \ref{fig:fig08}). The increase in positive time lag with energy suggests that as seed photons are reprocessed for longer time in the corona, emergent radiations gain energy through up-scattering, resulting in longer delays. 

Furthermore, for type-C QPOs, up-scattered photons directly reach the observer in low-inclination sources. For high-inclination sources, the initially large corona produces up-scattered X-ray photons through inverse Compton scattering. Most of these up-scattered photons escape without interacting significantly with the disc and leading to positive time lags. As the corona shrinks, a greater fraction of hard photons are redirected back toward the disc, where they interact with the thermalized disc, lose energy, and contribute to the softer X-ray emissions. This reprocessing mechanism eventually gives rise to negative lags. During the LHS and HIMS, the non-thermal flux fraction ($F_{\rm nth}$) is high ($\gtrsim 0.8$), indicating a large coronal size in both low- and high-inclination systems. In low-inclination sources, as the corona begins to shrink during HIMS ($F_{\rm nth} \sim 0.65-0.7$), the lag remains positive. Under similar conditions, however, high-inclination sources begin to show negative lags, suggesting more efficient reprocessing due to the increased interaction between coronal photons and the disc. When $F_{\rm nth}$ decreases further ($\lesssim 0.5$) and the corona becomes more radially compact, both low- and high-inclination sources exhibit negative lags. Interestingly, type-A QPOs are generally associated with minimal hard X-ray emissions compared to other QPOs and show a negative lag, except one observation in 2002 outburst of GX 339$-$4, suggesting that they originate from a smaller corona. In contrast, type-B QPOs, which appeared with Comptonized emissions similar to type-A QPOs but significantly lower than type-C QPOs, show positive lag for low inclination sources and both positive and negative lags for high inclination sources. With this, we argue that the corona associated with type-B QPOs is likely to differ in geometry from that of type-C QPOs, possibly being vertically elongated as it appears to be connected with radio ejections. This finding also corroborates the previous studies \cite{Homan-etal-2020,Belloni-etal-2020, Garcia-etal-2021,Mendez-etal-2022, Ma-etal-2023,Zhang-etal-2023}.

The present study indicates that type-A QPOs, which exhibit negative time lags, act as precursors to radio ejections in sources such as XTE J1752$-$223, XTE J1859$+$226, H1743$-$322, XTE J1550$-$564, and MAXI J1535$-$571, and may serve as potential indicators of changes in the coronal geometry. Notably, when type-B QPOs are observed with positive lags, a vertically elongated corona emerges leading to jet ejections in the form of radio flares \cite[see also][]{Stevens-Uttley-2016, Belloni-etal-2020, Kylafis-etal-2020, Chatterjee-etal-2020, Garcia-etal-2021, Harikrishna-Sriram-2022,Liu-etal-2022,Mendez-etal-2022, Zhang-etal-2023}. In contrast, type-B QPOs exhibiting negative lags are likely produced by a radially compact corona that is in the process of transitioning toward a vertical geometry, which possibly later results in jet ejection. The decrease in time lag of type-B QPOs with increasing radio and X-ray fluxes (Fig. \ref{fig:fig09}) for both low- and high-inclination sources further substantiates the scenario in which the radial extent of the corona diminishes as more coronal mass is redirected vertically, eventually leading to discrete jet ejection. This interpretation is further supported by the consistent disappearance of QPOs following such type-B events, although concurrent radio observations are not always available to confirm jet activity.

\subsection{Possible Disc Configurations with Source Inclination}

\begin{figure*}
    \begin{center}
    \includegraphics[width=\textwidth]{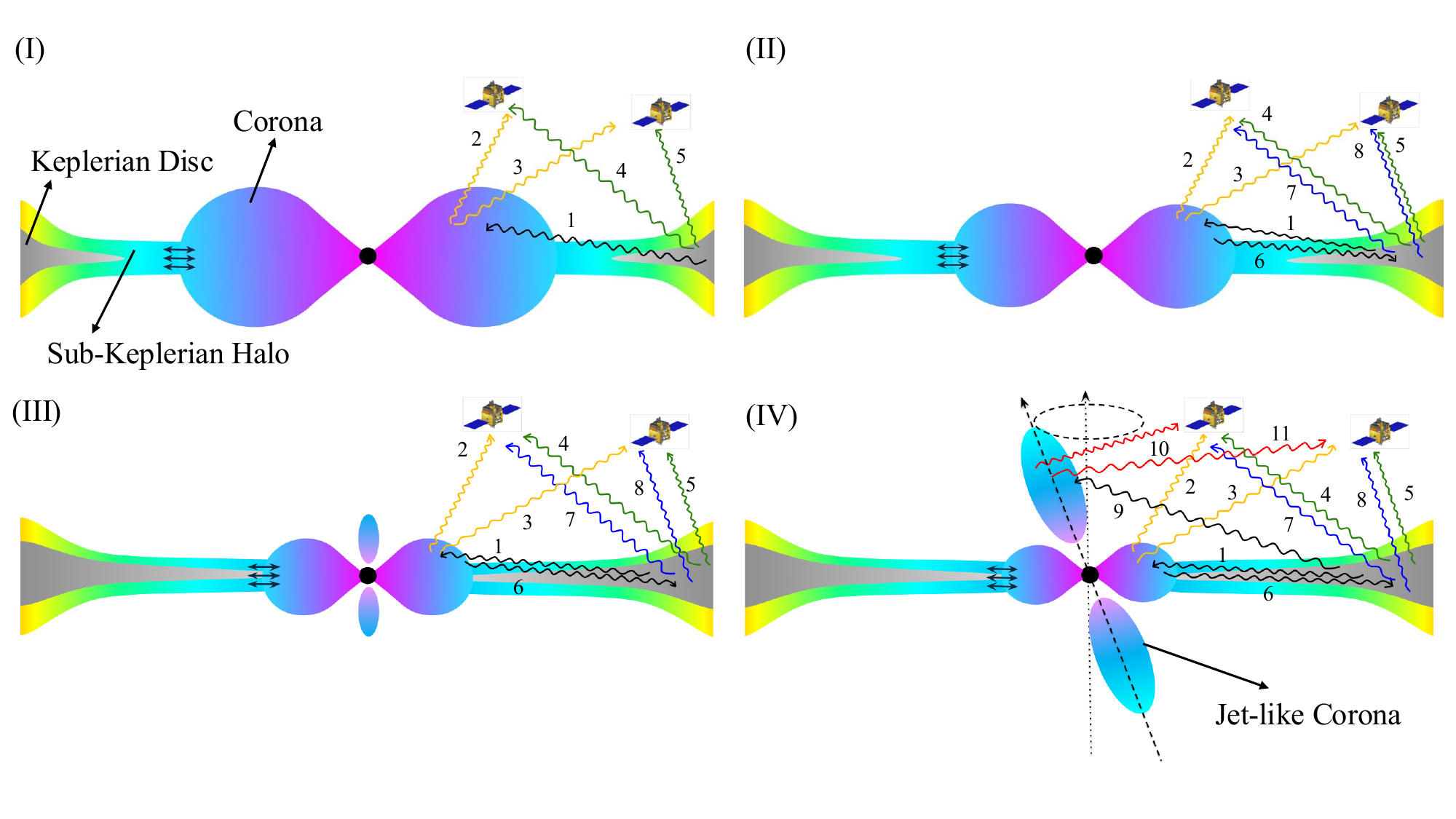}
    \end{center}
    \caption{Schematic representation of disc-jet scenarios: (I) Corona is radially extended and large in size, (II) Corona is radially extended but smaller in size, (III) Radial extent of corona is reduced and begins to elongate vertically, and (IV) Radial extent of corona is small and extends vertically. Soft photons reaching the observer directly from the disc are marked in green, while reprocessed photons are shown in blue. Hard photons reaching the observer from the radial corona is marked in yellow. Hard photons originating from the vertical corona, reaching the observer is shown in red. Soft photons incident on the corona from the disc and hard photons from corona to disc are shown in black. In (I), (II), (III) and (IV), the double headed arrows in the radial corona represent the oscillation of the corona causing type-C, type-A and type-B (negative lag) QPOs. In (IV), the dotted line marks the black hole’s spin axis, while the dashed line represents the misaligned jet axis, causing the precession. See the text for details.}
\label{fig:f12}
\end{figure*}

Following the discussions in the previous section, we explore the possible configuration of the disc-jet geometry in the context of time lag behavior and Comptonized emissions for type-A, type-B and type-C QPOs, as shown in Fig. \ref{fig:f12}. Four different possible disc-jet configurations are identified as delineated below. 

\begin{itemize}
    \item Configuration-I (Large Radially Extended Corona): A large radially extended corona leads to minimal reprocessing of up-scattered photons in the disc for both high and low-inclination sources resulting in positive lags for type-C QPOs with high normalized Comptonized flux.

    \item Configuration-II (Small Radially Extended Corona): Reduced coronal size increases the possibility of reprocessing of hard photons in the disc causing negative lags in high-inclination sources. The time lags remain positive in low-inclination sources as the effect of reprocessing remains weaker.

    \item Configuration-III (Radially and Vertically Compact Corona): As the radial extent of the corona shrinks and it begins to elongate vertically, reprocessing of hard photons in the disc increases regardless of the source inclination, resulting in negative lag. The modulation of the radially compact corona gives rise to type-A QPO, which seems to act as precursors to the subsequent jet ejections.

    \item Configuration-IV (Vertically Elongated Corona or Radially Compact Corona): The vertical elongation of the corona reduces the possibility of interaction between up-scattered photons and the disc, leading to a positive lag of type-B QPOs. This happens irrespective to the source inclinations. The presence of radio flares near type-B QPOs along with the reduced Comptonized X-ray flux further substantiates the concept of a vertically extended corona. When such a elongated corona precesses around the black hole's rotation axis, it could modulate the emerging hard radiation, leading to the formation of type-B QPOs. Moreover, in high-inclination sources, type-B QPOs with negative time lags possibly arise when up-scattered photons from the radially compact corona undergo additional reprocessing after interacting with the disc before reaching the observer.

\end{itemize}

We illustrate all four disc-jet configurations in Fig. \ref{fig:f12}, where the arrows marked with numbers represent the different photon paths. For example, (1) soft photons from the Keplerian disc incident upon the corona, (2) up-scattered photons reaching the observer at low-inclination, (3) up-scattered photons observed at high-inclination, (4) soft photons from the disc reaching the observer at low-inclination, (5) soft photons reaching the observer at high-inclination, (6) up-scattered photons returning to the disc, (7) reprocessed photons reaching the observer at low-inclination, and (8) reprocessed photons reaching the observer at high-inclination. (9) Soft photons interacting with the vertically extended corona, (10) up-scattered photons reaching a low-inclination observer from the vertically elongated corona, and (11) up-scattered photons reaching a high-inclination observer from the same vertically elongated corona. To distinguish between reprocessed soft photons and direct photons from the disc, we denote those reaching the observer in blue and green, respectively. The up-scattered photons reaching the observers from radially extended corona is shown in yellow, whereas photons reaching the observer from vertically elongated corona is shown in red. We show all other arrows in black.

\subsection{Jet velocity: Role of Mass Accretion and Spin}

We utilize the spectro-temporal properties of BH-XRBs under consideration to estimate their jet velocities. In doing so, we constrain the source spin and compute the disc radiative efficiency \cite[$\eta_{\rm acc}$,][]{Thorne-1974,Hobson-etal-2006}, while obtaining the jet velocity ($\beta$) as a function of $\varepsilon$. The details of the estimated $\beta$ and $\varepsilon$ for different observations across all sources are provided in Table \ref{tab:tab03}. 

We observe that $\beta$ generally increases with $\varepsilon$ (see Fig. \ref{fig:fig10}) and the estimated $\beta$ values (see Table \ref{tab:tab03}) are in good agreement with previously reported values measured from the proper motion of jet ejecta in different sources \cite[]{Corbel-etal-2005,Miller-etal-2011,Miller-etal-2012,Hannikainen-etal-2009}. We notice that in harder states (LHS and HIMS), jet velocity is sub-relativistic ($\lesssim 0.3\,c$) in nature, while in SIMS, it appears to be moderately relativistic ($\gtrsim 0.3-0.8\,c$) in presence of enhanced X-ray and radio luminosities, and sub-relativistic ($\lesssim 0.3\,c$) when luminosities are low. Furthermore, the correlation between intrinsic radio and X-ray luminosities (see Fig. \ref{fig:fig11}), indicating that higher mass accretion rates lead to more powerful jets. This possibly indicates a weak dependence between jet velocity and black hole spin, while showing a much stronger correlation with accretion rates, in line with earlier studies \cite[]{Fender-etal-2010, Russell-etal-2013, Fender-etal-2014, Aktar-etal-2015}. 

\begin{figure}
    \centering
    \includegraphics[width=\columnwidth]{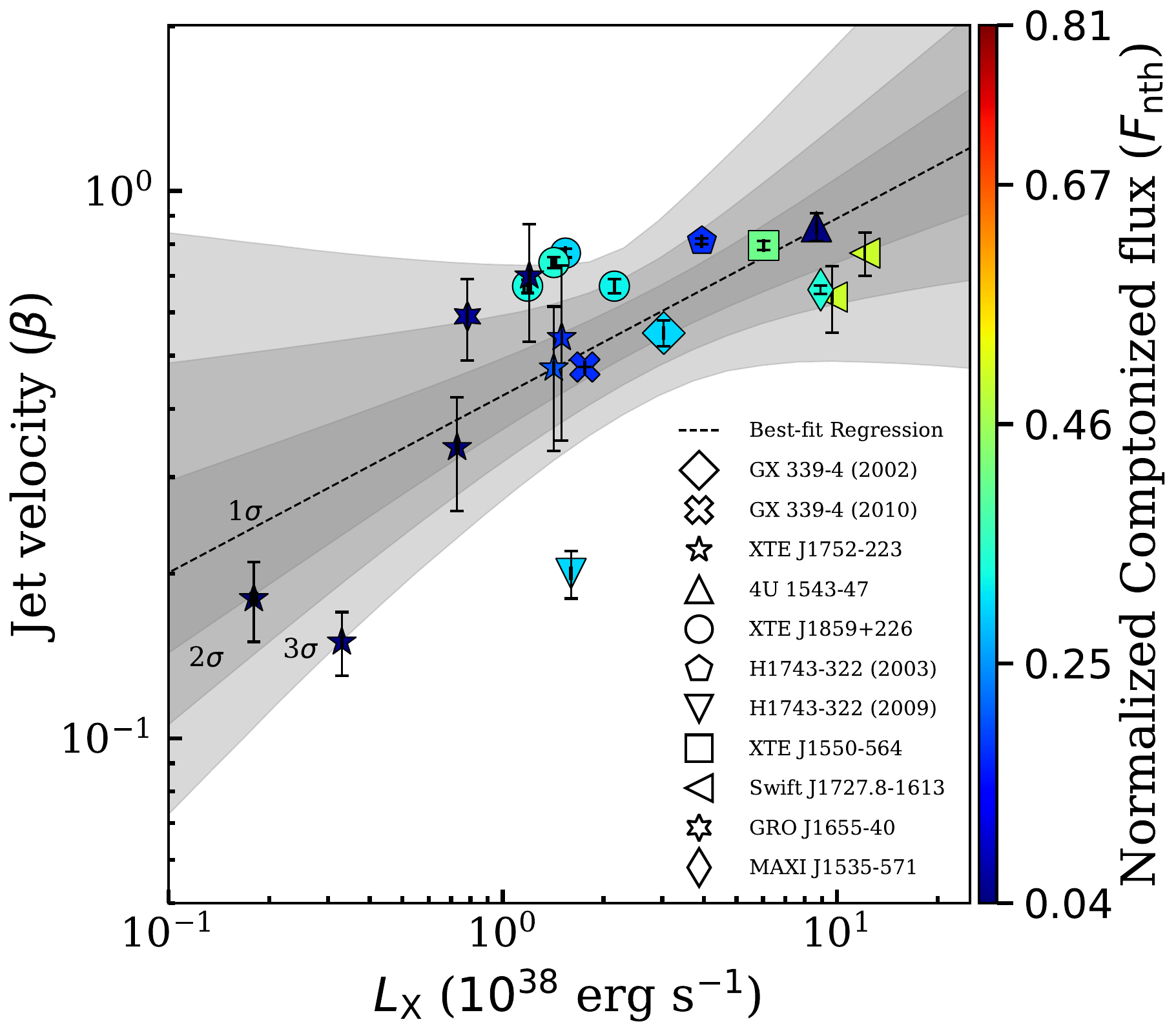}
    \caption{Correlation between X-ray luminosity and jet velocity. The dashed line represents the best-fit regression, while the shaded regions indicate the $1\sigma$, $2\sigma$ and $3\sigma$ confidence intervals. The color bar encodes values of $F_{\rm nth}$. Individual sources under consideration are marked using different symbols. See the text for details.}
    \label{fig:f13}
\end{figure}

Furthermore, we investigate the relationship between jet velocity ($\beta$) and bolometric X-ray luminosity ($L_{\rm X}$) during SIMS, where $L_{\rm X}$ is generally higher than in HIMS. Following the approach of \cite[]{Curran-2014}, we generate $10^6$ synthetic datasets using a bootstrap Monte Carlo method. Each dataset is constructed by sampling with replacement from the original $\beta$ and $L_{\rm X}$ values. To incorporate measurement uncertainties, each sampled point is perturbed using a normal distribution centered on the sampled value with a standard deviation equal to its observational error. The obtained results are presented in Fig. \ref{fig:f13}, where the best fitted dashed line corresponds to $\log \beta = (0.33 \pm 0.11) \log L_{\rm X} - (0.37 \pm 0.06)$. Different symbols represent individual sources, and the color bar denotes the corresponding $F_{\rm nth}$ values. Shaded regions indicate the $1\sigma$, $2\sigma$, and $3\sigma$ confidence intervals, with increasing transparency. The Pearson correlation coefficient $\sim 0.52$ indicates a moderate positive correlation, suggesting that higher X-ray luminosity (equivalently mass accretion rate) is associated with faster jet velocities ($\beta$) during SIMS.

Finally, we mention that the present formalism bears limitations. The Doppler correction is expected to depend on both inclination ($i$) and jet velocity ($\beta$). Moreover, the assumption that the changes in normalized Comptonized flux during radio flaring events directly correspond to $\varepsilon$ may be overly simplistic, as these flux variations could also reflect intrinsic changes in the accretion rate, rather than solely the outflow. Despite of all these, we argue that the basic conclusions of this work are expected to remain qualitatively unchanged.

We summarize the key findings of the present study in the context of disc-jet connection in BH-XRBs below:
\begin{enumerate}
    \item Strong Comptonized emissions with a radially extended corona appear to exhibit type-C QPOs in the BH-XRBs under consideration. Type-C QPOs show positive time lags for low-inclination sources, while negative time lags are observed for high-inclination sources, with the exception of Swift J1727.8$-$1613, XTE J1550$-$564, GRO J1655$-$40 and H1743$-$322 (2003 outburst).  
    
    \item Generally, type-A QPOs exhibit negative time lags with minimal Comptonized emissions regardless of source inclination angles. Our results indicate that type-A QPOs precede radio flares in several black hole X-ray binaries, including XTE J$1752-223$, XTE J$1859+226$, H$1743-322$, XTE J$1550-564$, and MAXI J$1535-571$. While this association points to a possible connection between type-A QPOs and jet ejection events, their infrequent detection calls for further observations to establish a robust relationship.

    \item Type-B QPOs are typically observed in the SIMS or during the spectral state transition from HIMS to SIMS with weak Comptonized emissions similar to type-A QPOs. We observed that type-B QPOs are strongly associated with radio flares and exhibit positive lag for low inclination sources and both positive and negative lags for high inclination sources. This strongly suggests that type-B QPOs are linked to the geometry of a vertically extended corona.

    \item Our findings demonstrate a strong correlation between the observed radio and X-ray luminosities, providing compelling evidence that jets are predominantly powered by accretion. We observe that during SIMS, the jet velocity appears to be moderately relativistic ($\gtrsim 0.3-0.8\,c$) when both X-ray and radio luminosities are high, and remains sub-relativistic ($\lesssim 0.3\,c$) when the luminosities are low for the BH-XRBs under consideration.
\end{enumerate}

\section{Acknowledgment}

Authors thank the reviewer for detailed and constructive comments, which help to improve the quality of the manuscript. SDC and SD thank the Department of Physics, IIT Guwahati, for providing the facilities to complete this work. SDC also acknowledges the financial support provided by Prime Minister Research Fellowship program to carry out this work. BGR thanks Gifu University, Japan for support. AN thanks GH, SAG; DD, PDMSA, and Director, URSC for encouragement and continuous support to carry out this research. This research has made use of data obtained from High Energy Astrophysics Archive Research Center (HEASARC) online services, provided by NASA/Goddard Space Flight Center. 

\section{Data Availability}

The data utilized in this work is available in the websites of {\url{https://heasarc.gsfc.nasa.gov/cgi-bin/W3Browse/w3browse.pl}} and \url{https://archive.hxmt.cn/proposal} for \textit{RXTE} and \textit{HXMT}, respectively.

\input{ms_01.bbl}

\end{document}